\pdfoutput=1
\documentclass{moriond}

\def\be{\begin{equation}}
\def\ee{\end{equation}}
\def\bea{\begin{eqnarray}}
\def\eea{\end{eqnarray}}

\def\thetal {\ensuremath{\theta_{l}}\xspace}
\def\thetak {\ensuremath{\theta_{K}}\xspace}

\usepackage{ifthen} % for conditional statements
\usepackage{xspace} % To avoid problems with missing or double spaces after
\usepackage{upgreek} % Adds in support for greek letters in roman typeset
\newboolean{uprightparticles}
\setboolean{uprightparticles}{false} %True for upright particle symbols
\def\MagUp {\mbox{\em Mag\kern -0.05em Up}\xspace}

\ifthenelse{\boolean{uprightparticles}}%
{

 \def\Pmu         {\ensuremath{\upmu}\xspace}

 \def\Ppi         {\ensuremath{\uppi}\xspace}                 
                  
 \def\Prho        {\ensuremath{\uprho}\xspace}

 \def\Ppsi        {\ensuremath{\uppsi}\xspace}

 \def\PDelta      {\ensuremath{\Delta}\xspace}                 
 \def\PXi      {\ensuremath{\Xi}\xspace}                 
 \def\PLambda      {\ensuremath{\Lambda}\xspace}                 
 \def\PSigma      {\ensuremath{\Sigma}\xspace}                 
 \def\POmega      {\ensuremath{\Omega}\xspace}                 
 \def\PUpsilon      {\ensuremath{\Upsilon}\xspace}                 
 
 \def\PB      {\ensuremath{\mathrm{B}}\xspace}                 
                  
 \def\PD      {\ensuremath{\mathrm{D}}\xspace}

 \def\PJ      {\ensuremath{\mathrm{J}}\xspace}                 
 \def\PK      {\ensuremath{\mathrm{K}}\xspace}

 \def\Pb      {\ensuremath{\mathrm{b}}\xspace}

 \def\Pe      {\ensuremath{\mathrm{e}}\xspace}

 \def\Pi      {\ensuremath{\mathrm{i}}\xspace}

 \def\Ps      {\ensuremath{\mathrm{s}}\xspace}

}
{

 \def\Pmu         {\ensuremath{\mu}\xspace}

 \def\Ppi         {\ensuremath{\pi}\xspace}                 
                  
 \def\Prho        {\ensuremath{\rho}\xspace}

 \def\Ppsi        {\ensuremath{\psi}\xspace}                 
                  
 \mathchardef\PDelta="7101
 \mathchardef\PXi="7104
 \mathchardef\PLambda="7103
 \mathchardef\PSigma="7106
 \mathchardef\POmega="710A
 \mathchardef\PUpsilon="7107
                  
 \def\PB      {\ensuremath{B}\xspace}                 
                  
 \def\PD      {\ensuremath{D}\xspace}

 \def\PJ      {\ensuremath{J}\xspace}                 
 \def\PK      {\ensuremath{K}\xspace}

 \def\Pb      {\ensuremath{b}\xspace}

 \def\Pe      {\ensuremath{e}\xspace}

 \def\Pi      {\ensuremath{i}\xspace}

 \def\Ps      {\ensuremath{s}\xspace}

}

\makeatletter
\ifcase \@ptsize \relax% 10pt
  \newcommand{\miniscule}{\@setfontsize\miniscule{4}{5}}% \tiny: 5/6
\or% 11pt
  \newcommand{\miniscule}{\@setfontsize\miniscule{5}{6}}% \tiny: 6/7
\or% 12pt
  \newcommand{\miniscule}{\@setfontsize\miniscule{5}{6}}% \tiny: 6/7
\fi
\makeatother

\DeclareRobustCommand{\optbar}[1]{\shortstack{{\miniscule (\rule[.5ex]{1.25em}{.18mm})}
  \\ [-.7ex] $#1$}}

\def\en         {{\ensuremath{\Pe^-}}\xspace}   % electron negative (\em is taken)
\def\ep         {{\ensuremath{\Pe^+}}\xspace}

\def\mup        {{\ensuremath{\Pmu^+}}\xspace}
\def\mun        {{\ensuremath{\Pmu^-}}\xspace} % muon negative (\mum is taken)
\def\mumu       {{\ensuremath{\Pmu^+\Pmu^-}}\xspace}

\def\squark    {{\ensuremath{\Ps}}\xspace}

\def\bquark    {{\ensuremath{\Pb}}\xspace}

\def\pion   {{\ensuremath{\Ppi}}\xspace}
\def\piz    {{\ensuremath{\pion^0}}\xspace}

\def\pip    {{\ensuremath{\pion^+}}\xspace}
\def\pim    {{\ensuremath{\pion^-}}\xspace}

\def\rhomeson {{\ensuremath{\Prho}}\xspace}
\def\rhoz     {{\ensuremath{\rhomeson^0}}\xspace}

\def\kaon    {{\ensuremath{\PK}}\xspace}
  \def\Kbar    {{\kern 0.2em\overline{\kern -0.2em \PK}{}}\xspace}

\def\KorKbar    {\kern 0.18em\optbar{\kern -0.18em K}{}\xspace}

\def\Kp      {{\ensuremath{\kaon^+}}\xspace}

\def\KS      {{\ensuremath{\kaon^0_{\rm\scriptscriptstyle S}}}\xspace}

\def\Kstarz  {{\ensuremath{\kaon^{*0}}}\xspace}

\def\Kstarp  {{\ensuremath{\kaon^{*+}}}\xspace}

  \def\Dbar    {{\kern 0.2em\overline{\kern -0.2em \PD}{}}\xspace}

\def\DorDbar    {\kern 0.18em\optbar{\kern -0.18em D}{}\xspace}

\def\B       {{\ensuremath{\PB}}\xspace}
\def\Bbar    {{\ensuremath{\kern 0.18em\overline{\kern -0.18em \PB}{}}}\xspace}

\def\BorBbar    {\kern 0.18em\optbar{\kern -0.18em B}{}\xspace}

\def\Bu      {{\ensuremath{\B^+}}\xspace}

\def\Bd      {{\ensuremath{\B^0}}\xspace}
\def\Bs      {{\ensuremath{\B^0_\squark}}\xspace}

\def\jpsi     {{\ensuremath{{\PJ\mskip -3mu/\mskip -2mu\Ppsi\mskip 2mu}}}\xspace}
\def\psitwos  {{\ensuremath{\Ppsi{(2S)}}}\xspace}

  \def\Y#1S{\ensuremath{\PUpsilon{(#1S)}}\xspace}% no space before {...}!

\def\Lz          {{\ensuremath{\PLambda}}\xspace}
\def\Lbar        {{\ensuremath{\kern 0.1em\overline{\kern -0.1em\PLambda}}}\xspace}
\def\LorLbar    {\kern 0.18em\optbar{\kern -0.18em \PLambda}{}\xspace}

\def\Lb      {{\ensuremath{\Lz^0_\bquark}}\xspace}

\newcommand{\decay}[2]{\ensuremath{#1\!\to #2}\xspace}         % {\Pa}{\Pb \Pc}

\def\to                 {\ensuremath{\rightarrow}\xspace}

\def\CP                {{\ensuremath{C\!P}}\xspace}

\def\AT#1     {\ensuremath{A_{\mathrm{T}}^{#1}}\xspace}           % 2

\def\C#1      {\ensuremath{\mathcal{C}_{#1}}\xspace}                       % 9
\def\Cp#1     {\ensuremath{\mathcal{C}_{#1}^{'}}\xspace}                    % 7
\def\Ceff#1   {\ensuremath{\mathcal{C}_{#1}^{\mathrm{(eff)}}}\xspace}        % 9  
\def\Cpeff#1  {\ensuremath{\mathcal{C}_{#1}^{'\mathrm{(eff)}}}\xspace}       % 7
\def\Ope#1    {\ensuremath{\mathcal{O}_{#1}}\xspace}                       % 2
\def\Opep#1   {\ensuremath{\mathcal{O}_{#1}^{'}}\xspace}                    % 7

\newcommand{\tev}{\ifthenelse{\boolean{inbibliography}}{\ensuremath{~T\kern -0.05em eV}\xspace}{\ensuremath{\mathrm{\,Te\kern -0.1em V}}}\xspace}
\newcommand{\gev}{\ensuremath{\mathrm{\,Ge\kern -0.1em V}}\xspace}
\newcommand{\mev}{\ensuremath{\mathrm{\,Me\kern -0.1em V}}\xspace}
\newcommand{\kev}{\ensuremath{\mathrm{\,ke\kern -0.1em V}}\xspace}
\newcommand{\ev}{\ensuremath{\mathrm{\,e\kern -0.1em V}}\xspace}
\newcommand{\gevc}{\ensuremath{{\mathrm{\,Ge\kern -0.1em V\!/}c}}\xspace}
\newcommand{\mevc}{\ensuremath{{\mathrm{\,Me\kern -0.1em V\!/}c}}\xspace}
\newcommand{\gevcc}{\ensuremath{{\mathrm{\,Ge\kern -0.1em V\!/}c^2}}\xspace}
\newcommand{\gevgevcccc}{\ensuremath{{\mathrm{\,Ge\kern -0.1em V^2\!/}c^4}}\xspace}
\newcommand{\mevcc}{\ensuremath{{\mathrm{\,Me\kern -0.1em V\!/}c^2}}\xspace}

\def\invfb   {\ensuremath{\mbox{\,fb}^{-1}}\xspace}

\def\deriv {\ensuremath{\mathrm{d}}}

\def\gsim{{~\raise.15em\hbox{$>$}\kern-.85em
          \lower.35em\hbox{$\sim$}~}\xspace}
\def\lsim{{~\raise.15em\hbox{$<$}\kern-.85em
          \lower.35em\hbox{$\sim$}~}\xspace}

\def\tell1  {TELL1\xspace}
\def\ukl1   {UKL1\xspace}

\newcommand{\vs}{\mbox{\itshape vs.}\xspace}

\newcommand{\Photo}{}

\begin{document}
\vspace*{4cm}
\title{Latest results on rare decays from LHCb}

\author{C.\ Langenbruch on behalf of the LHCb collaboration}

\address{University of Warwick, 
  Department of Physics,\\
  Gibbet Hill Road,
  Coventry CV4 7AL, UK
}

\maketitle\abstracts{
  Rare flavour changing neutral current decays are sensitive indirect probes for new effects beyond the Standard Model (SM). 
  In the SM, these decays are forbidden at tree level and are therefore loop-suppressed.
  In SM extensions, new, heavy particles can significantly contribute and
  affect both their branching fractions as well as their angular distributions.\\
  The rare decay $B^0\to K^{*0}(\to K^+\pi^-)\mu^+\mu^-$ 
  is of particular interest, since it gives access to many angular observables,
  allowing to model-independently test the operator structure of the decay. 
  A previous analysis of the angular distributions of the final state particles showed interesting tensions with SM predictions
  using the data sample taken by the LHCb detector during 2011. 
  These proceedings will summarize latest results on rare decays from the LHCb experiment 
  with emphasis on the angular analysis of the decay $B^0\to K^{*0}\mu^+\mu^-$,  
  using the full Run I data sample of the LHCb experiment.
}

\section{Introduction}
Rare flavour changing neutral current (FCNC) decays 
are,
in the Standard Model (SM), forbidden at lowest perturbative order and proceed via loop-order diagrams. 
New heavy particles in extensions of the SM can appear in competing Feynman diagrams and significantly affect
both the branching fractions of rare decays and the angular distributions of the final state particles.
Studies of rare decays therefore constitute sensitive searches for effects beyond the SM, 
and furthermore allow to probe the underlying operator structure via global fits~\cite{Descotes-Genon:2013wba,Beaujean:2013soa,Altmannshofer:2014rta,Hurth:2013ssa}.

\section{Angular analysis of the rare decay $\decay{\Bd}{\Kstarz\mumu}$}
\subsection{Angular observables in $\decay{\Bd}{\Kstarz\mumu}$}
The rare decay $\decay{\Bd}{\Kstarz\mumu}$ is of particular interest since the $\Kp\pim\mumu$ final state allows access to many angular observables.
The final state is fully defined by the three decay angles $\vec{\Omega}=(\cos\thetal,\cos\thetak,\phi)$, and $q^2$, the invariant mass of the dilepton system squared~\cite{Altmannshofer:2008dz}. 
The \CP-averaged angular distribution of the decay $\decay{\Bd}{\Kstarz\mumu}$ in a bin of $q^2$ is given by
\begin{eqnarray}
\frac{1}{\deriv(\Gamma+\bar{\Gamma})/\deriv q^2}\frac{\deriv^3(\Gamma+\bar{\Gamma})}{\deriv\vec{\Omega}} =
\frac{9}{32\pi} &\Big[
 & \frac{3}{4} (1-{F_{\rm L}})\sin^2\thetak + {F_{\rm L}}\cos^2\thetak  \label{eq:pdfpwave}
+ \frac{1}{4}(1-{F_{\rm L}})\sin^2\thetak\cos 2\thetal\nonumber\\
&-& {F_{\rm L}} \cos^2\thetak\cos 2\thetal + {S_3}\sin^2\thetak \sin^2\thetal \cos 2\phi\nonumber\\
&+& {S_4} \sin 2\thetak \sin 2\thetal \cos\phi + {S_5}\sin 2\thetak \sin \thetal \cos \phi\nonumber\\
&+& \frac{4}{3} {A_{\rm FB}} \sin^2\thetak \cos\thetal + {S_7} \sin 2\thetak \sin\thetal \sin\phi\nonumber\\
&+& {S_8} \sin 2\thetak \sin 2\thetal \sin\phi + {S_9}\sin^2\thetak \sin^2\thetal \sin 2\phi  \Big].
\end{eqnarray}
Here, $F_{\rm L}$ denotes the longitudinal polarization fraction of the $\Kstarz$ and $A_{\rm FB}$ the forward-backward asymmetry of the dimuon system. 
The LHCb collaboration performed two angular analyses~\cite{Aaij:2013iag,Aaij:2013qta} using angular folding techniques to determine the observables with the data taken during 2011, which corresponds to an integrated luminosity of $1\invfb$. 
While all the angular observables in Ref.~\cite{Aaij:2013iag} are found to be in good agreement with SM predictions,
the measurement of the less form-factor dependent observable $P_5^\prime=S_5/\sqrt{F_{\rm L}(1-F_{\rm L})}$, that was proposed in Ref.~\cite{Descotes-Genon:2013vna}, 
shows a local deviation from the SM prediction, corresponding to $3.7$ standard deviations ($\sigma$)~\cite{Aaij:2013qta}. 
Global fits show that the tension can be reduced by a negative shift of the Wilson coefficient ${\cal C}_9$ which parametrises the vector coupling strength~\cite{Descotes-Genon:2013wba,Beaujean:2013soa,Altmannshofer:2014rta,Hurth:2013ssa}. 
New physics (NP) explanations for this shift include the possibility of heavy $Z^\prime$ gauge bosons~\cite{Gauld:2013qba,Buras:2013qja,Altmannshofer:2013foa,Glashow:2014iga,Crivellin:2015mga} or leptoquarks~\cite{Hiller:2014yaa,Biswas:2014gga,Buras:2014fpa,Gripaios:2014tna}.
However, the significance of the tension could be reduced if hadronic uncertainties are underestimated~\cite{Jager:2012uw,Jager:2014rwa,Lyon:2014hpa}. 

\subsection{Selection of signal candidates}
\begin{figure}
  \centering
  \includegraphics[height=3.75cm,clip=true,trim=0mm 0mm 0mm 10mm]{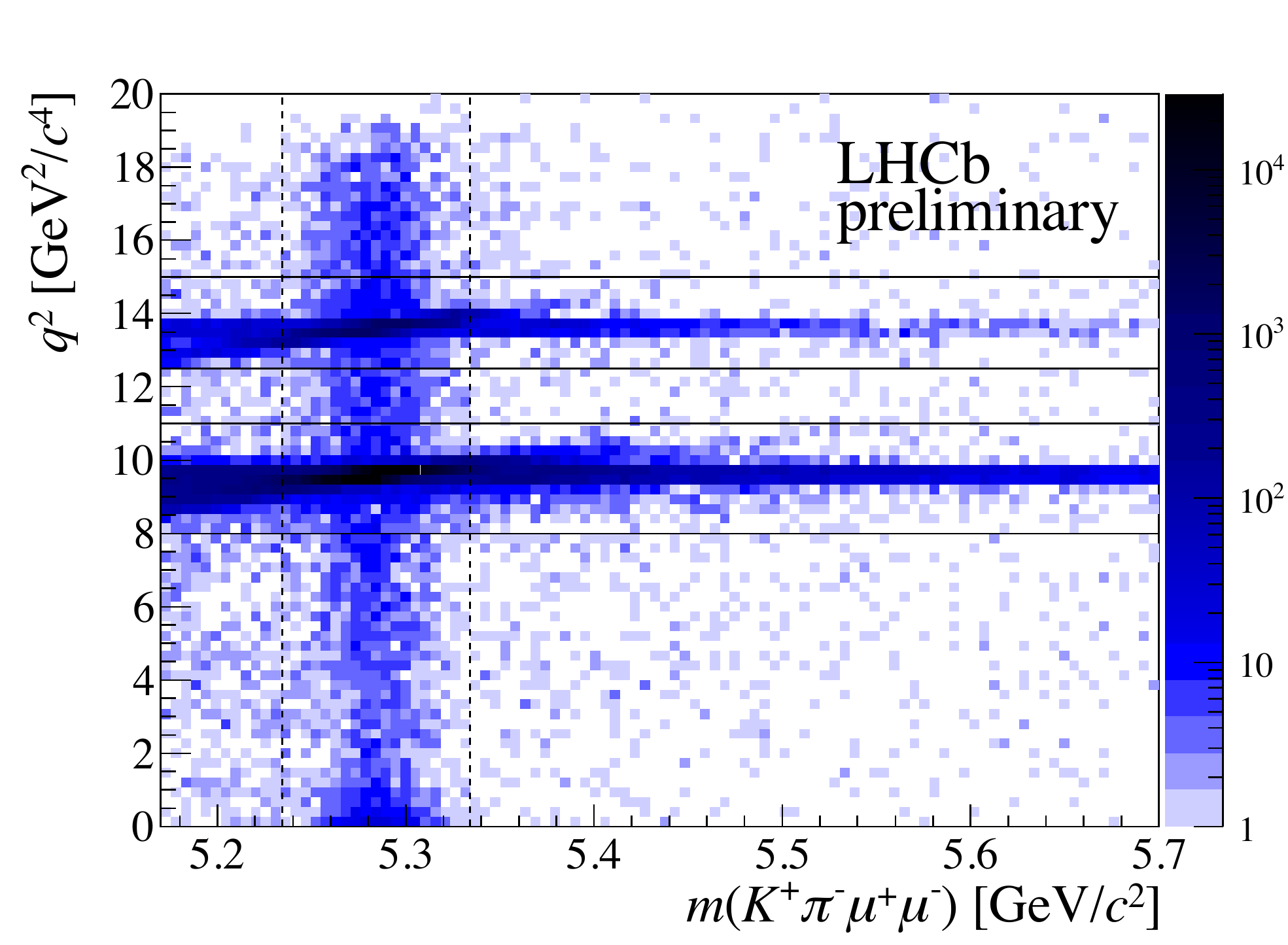}
  \includegraphics[height=3.75cm]{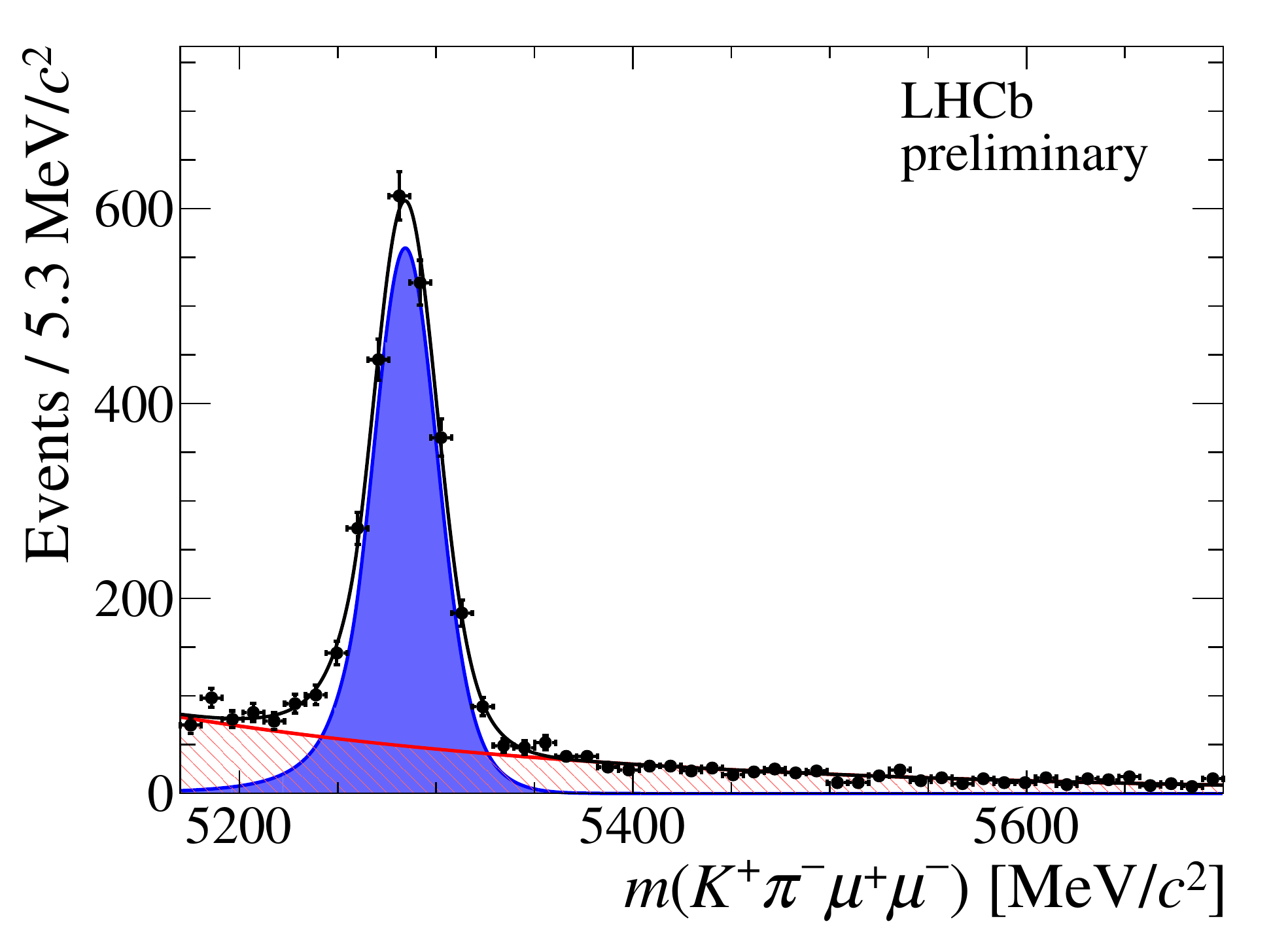}
  \includegraphics[height=3.75cm]{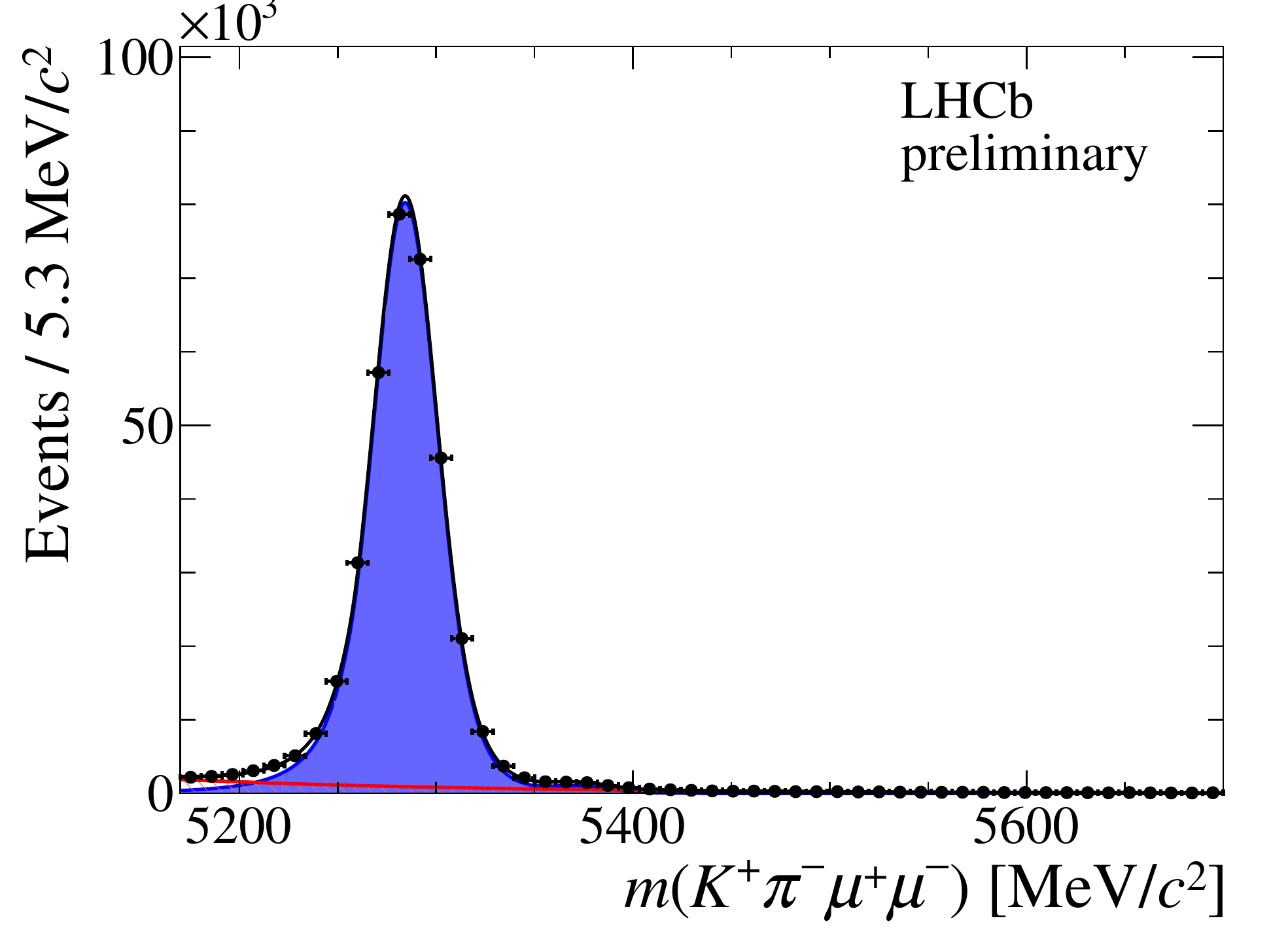}
  \caption{
    (Left) Invariant mass of the $\Kp\pim\mumu$ system \vs\ $q^2$.
    The signal decay $\decay{\Bd}{\Kstarz\mumu}$ is clearly visible as vertical band.
    (Middle) The signal decay $\decay{\Bd}{\Kstarz\mumu}$ integrated over $q^2$.
    The signal yield is $2\,398\pm 57$.
    (Right) The high-statistics control mode $\decay{\Bd}{\jpsi\Kstarz}$.
    \label{fig:mbvsq2}}
\end{figure}

An update of the angular analysis of $\decay{\Bd}{\Kstarz\mumu}$ using the full Run I data sample corresponding to $3\invfb$ was eagerly awaited in the community
and preliminary results are presented here for the first time~\cite{LHCb-CONF-2015-002}. 
The selection of $\decay{\Bd}{\Kstarz\mumu}$ signal candidates is improved compared to Ref.~\cite{Aaij:2013iag,Aaij:2013qta} with a simplified, yet more efficient, multivariate classifier 
to reduce combinatorial background events and more stringent vetoes to reject peaking backgrounds. 
Figure~\ref{fig:mbvsq2} gives the distribution of the invariant mass of the $\Kp\pim\mumu$ system \vs\ $q^2$ for signal candidates after the full selection,
where the $\decay{\Bd}{\Kstarz\mumu}$ signal decay is clearly visible as vertical band. 
The $q^2$ regions $8.0<q^2<11.0\gevgevcccc$ and $12.5<q^2<15.0\gevgevcccc$ contain
the tree-level decays $\decay{\Bd}{\jpsi(\to\mup\mun)\Kstarz}$ and $\decay{\Bd}{\psitwos(\to\mup\mun)\Kstarz}$ 
which are used as important control decays but
vetoed when selecting $\decay{\Bd}{\Kstarz\mumu}$ signal candidates. 
Integrated over $q^2$, the $\decay{\Bd}{\Kstarz\mumu}$ signal yield is $2\,398\pm 57$, as shown in Fig.~\ref{fig:mbvsq2}. 

\subsection{Angular analysis}
The angular observables are determined by performing an unbinned maximum likelihood fit 
in bins of $q^2$, 
using a $q^2$ binning which is both finer and more regular than the binning used in Refs.~\cite{Aaij:2013iag,Aaij:2013qta}.
The fit uses the invariant mass distribution of the $\Kp\pim\mumu$ system, the invariant $\Kp\pim$ mass and
the three-decay angles as input distributions without applying angular foldings.
This allows to quote the covariance matrices for all eight angular observables which is important for the use of the results in global fits. 
The invariant $\Kp\pim$ mass distribution is used to constrain the contribution from 
events where the $\Kp\pim$ system is in a spin-0 configuration, the so-called S-wave. 
The additional six parameters for the description of the S-wave and the interference terms with the
$\Kstarz$ P-wave are treated as nuisance parameters and allowed to vary in the fit. 
The trigger, reconstruction and selection of signal events causes distortions of the distributions of $q^2$ and the decay angles. 
This acceptance effect is modelled using a multidimensional combination of Legendre polynomials.  
The polynomial coefficients are determined using a moments analysis of a large sample of simulated $\decay{\Bd}{\Kstarz\mumu}$ events,
generated according to a phase-space model. 
The Feldman-Cousins method~\cite{1998PhRvD..57.3873F} is used to guarantee correct coverage for the angular observables even for low signal yields. 

\subsection{Results}
\begin{figure}
  \centering
  \includegraphics[width=0.48\linewidth]{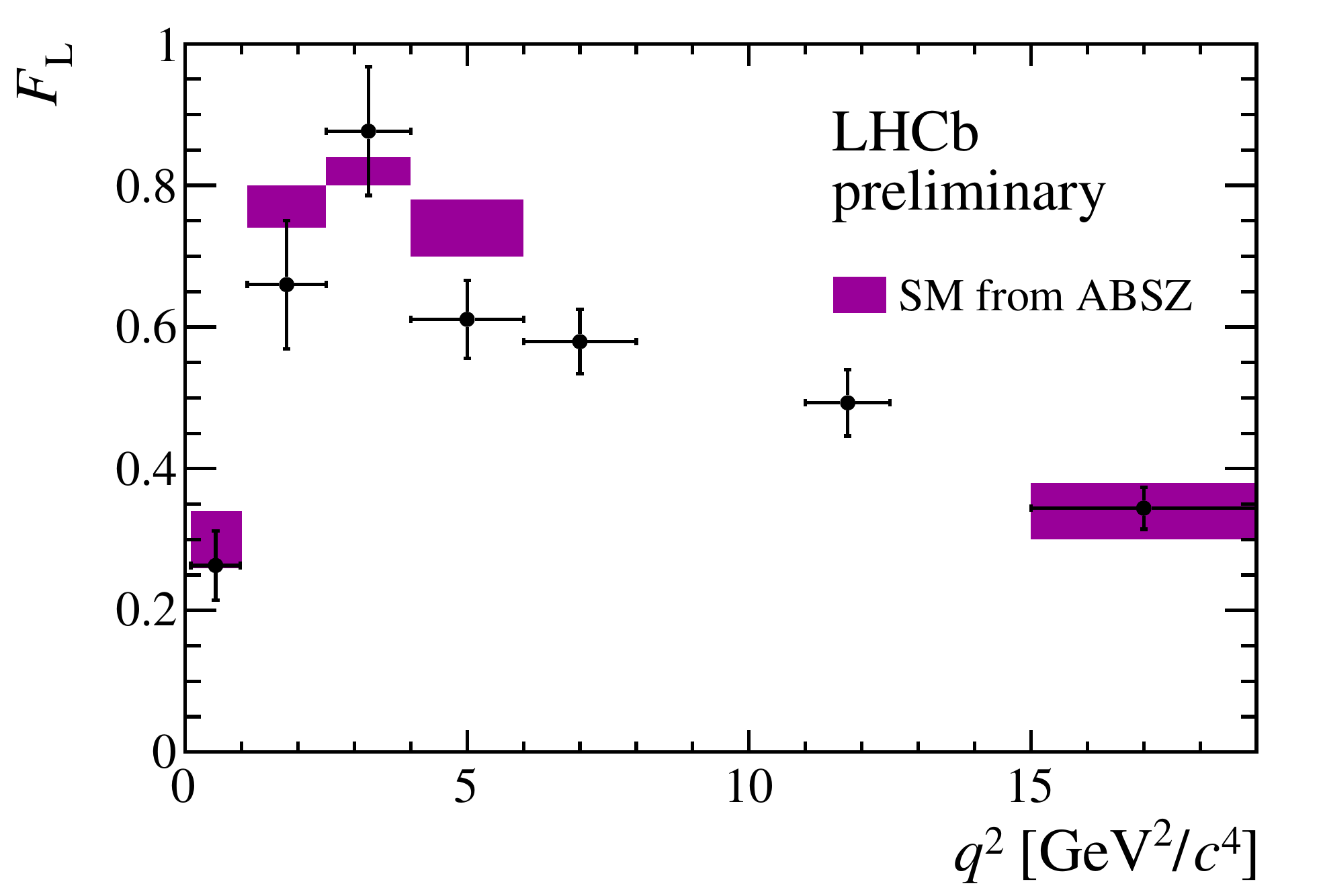} 
  \includegraphics[width=0.48\linewidth]{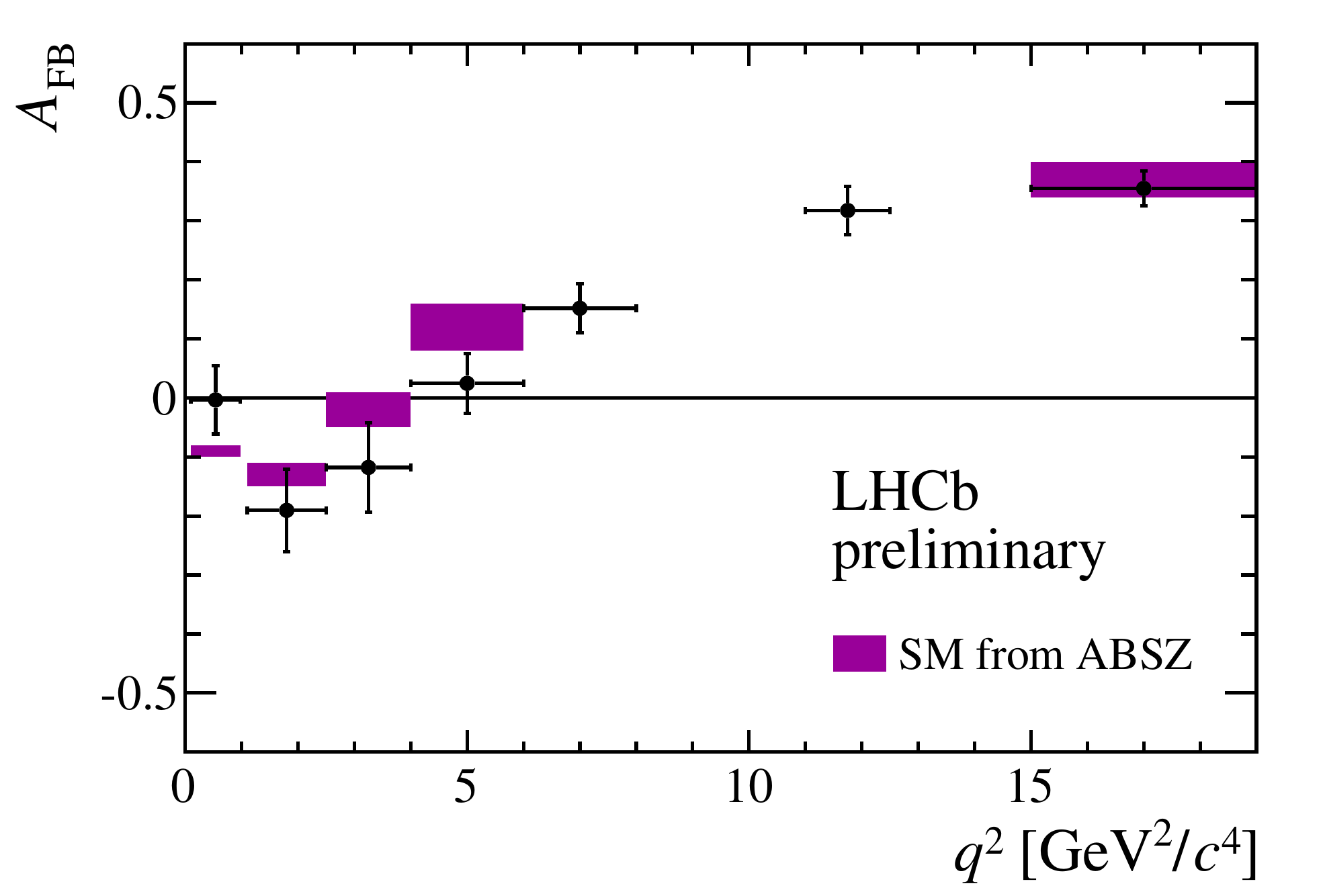} \\
  \includegraphics[width=0.48\linewidth]{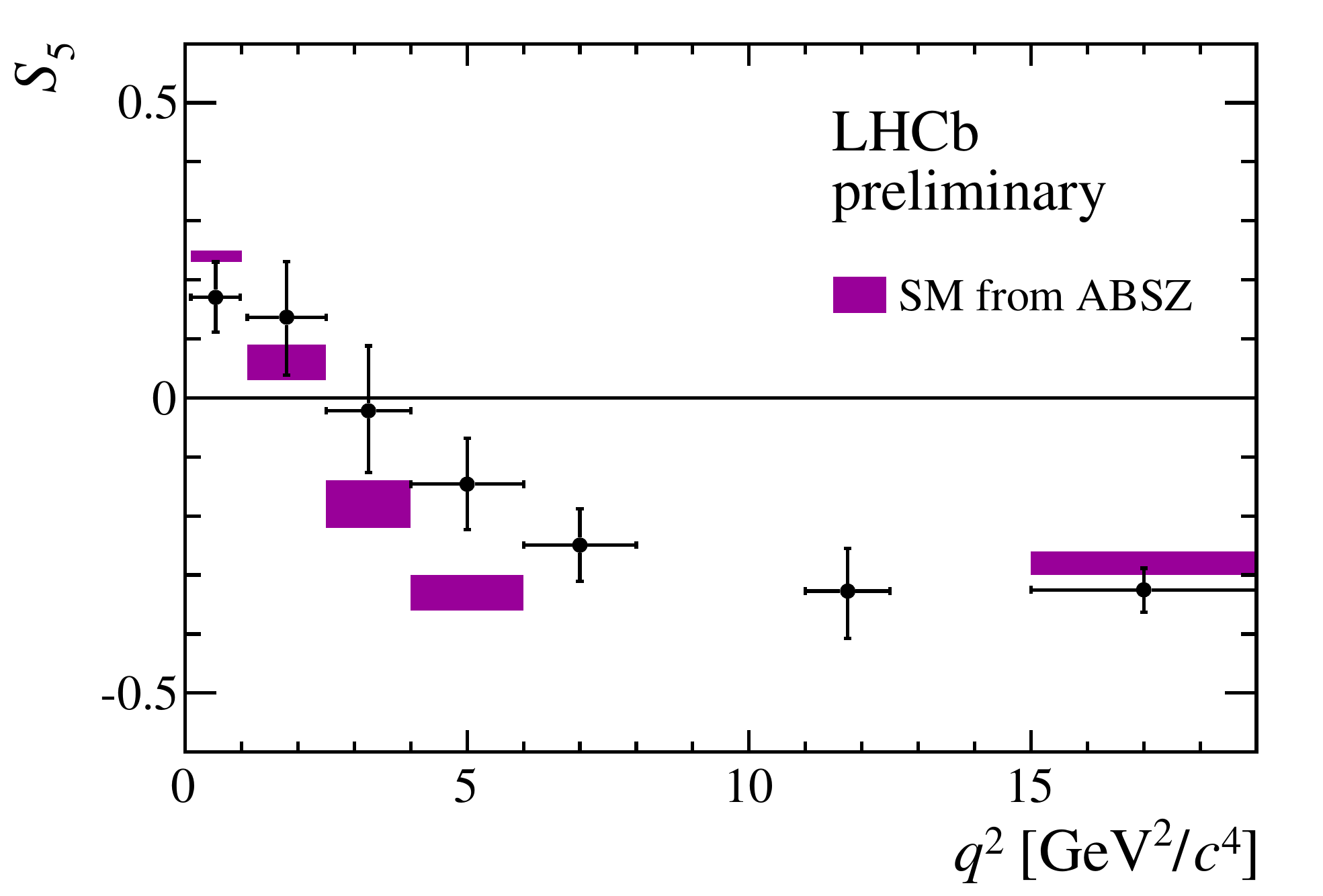} 
  \includegraphics[width=0.48\linewidth]{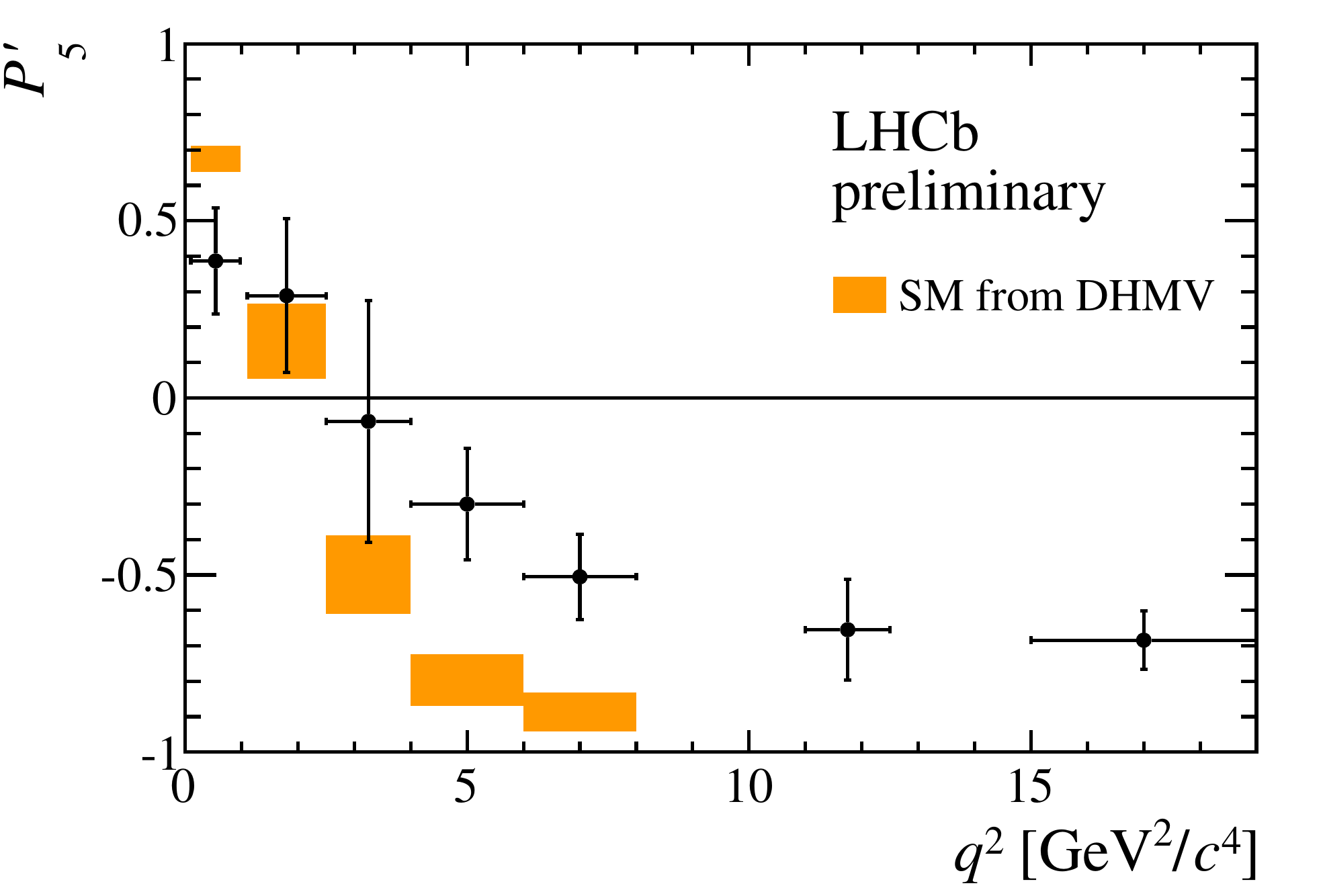}
  \caption{The angular observables $F_{\rm L}$, $A_{\rm FB}$, $S_5$ and $P_5^\prime$ overlaid with SM predictions from (purple) Ref.~\protect\cite{Altmannshofer:2014rta,Straub:2015ica} and (orange) Ref.~\protect\cite{Descotes-Genon:2014uoa}.\label{fig:angularobs}}
\end{figure}

The results for $F_{\rm L}$, $A_{\rm FB}$, $S_5$ and $P_5^\prime$ are given in Fig.~\ref{fig:angularobs},
overlaid with SM predictions from Refs.~\cite{Altmannshofer:2014rta,Straub:2015ica} and Ref.~\cite{Descotes-Genon:2014uoa}. 
The longitudinal polarization fraction $F_{\rm L}$ and the forward-backward asymmetry $A_{\rm FB}$ are found to be in good agreement with SM predictions. 
Interestingly, for $A_{\rm FB}$, the data points seem to lie systematically below the predictions in the $1.1<q^2<6\gevgevcccc$ range. 
The measurement of the less form-factor dependent observable $P_5^\prime$ is found to be compatible with the previous publication~\cite{Aaij:2013qta} 
and lies above the SM prediction~\cite{Descotes-Genon:2014uoa} in the $4.0<q^2<8.0\gevgevcccc$ region. 
The deviation from the SM prediction 
corresponds to $2.9\,\sigma$ for each of the two $q^2$ bins in the region $4.0<q^2<6.0\gevgevcccc$ and $6.0<q^2<8.0\gevgevcccc$. 
Neglecting correlations between the bins, 
the $\chi^2$ probability to find a deviation of this size or larger for two degrees of freedom  
results in a naive significance of $3.7\,\sigma$. 
The remaining observables $S_3$, $S_4$, $S_7$, $S_8$ and $S_9$ are given in Ref.~\cite{LHCb-CONF-2015-002} and show good agreement with SM predictions. 

\section{Branching fraction measurements of $\decay{B}{K^{(*)}\mumu}$ and $\Bs\to\phi\mumu$ decays}
\begin{figure}
  \centering
  \includegraphics[width=0.32\linewidth]{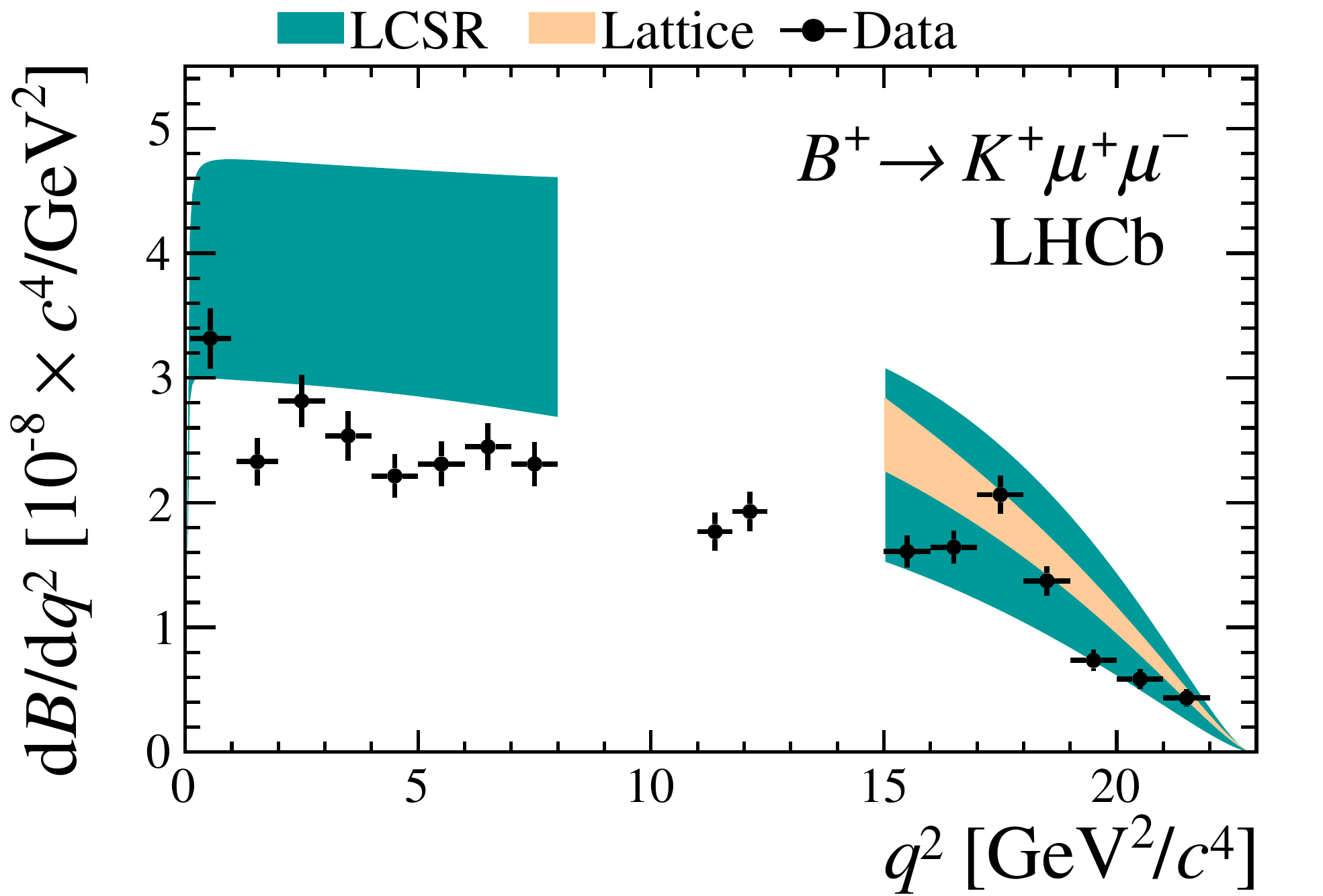} 
  \includegraphics[width=0.32\linewidth]{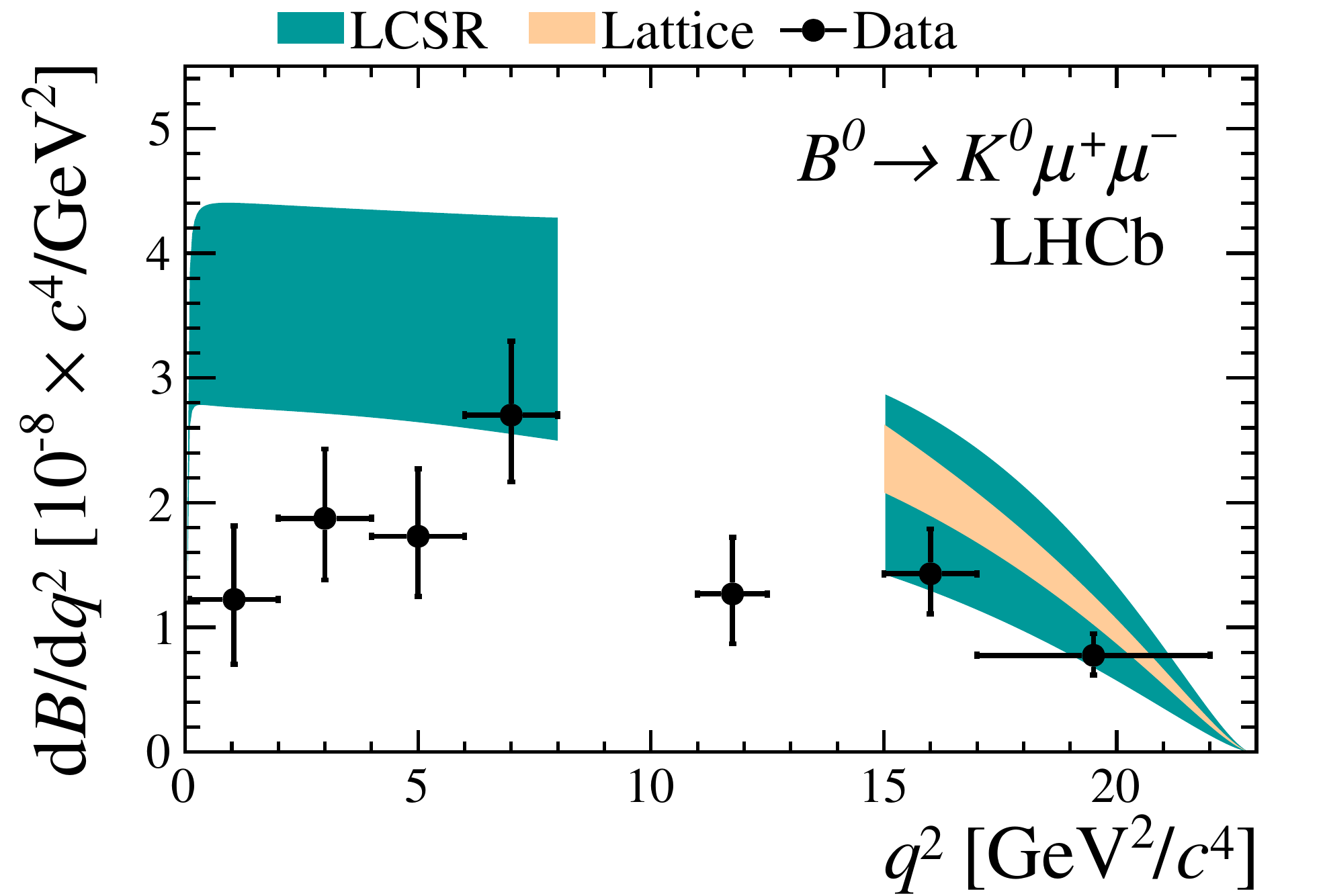}
  \includegraphics[width=0.32\linewidth]{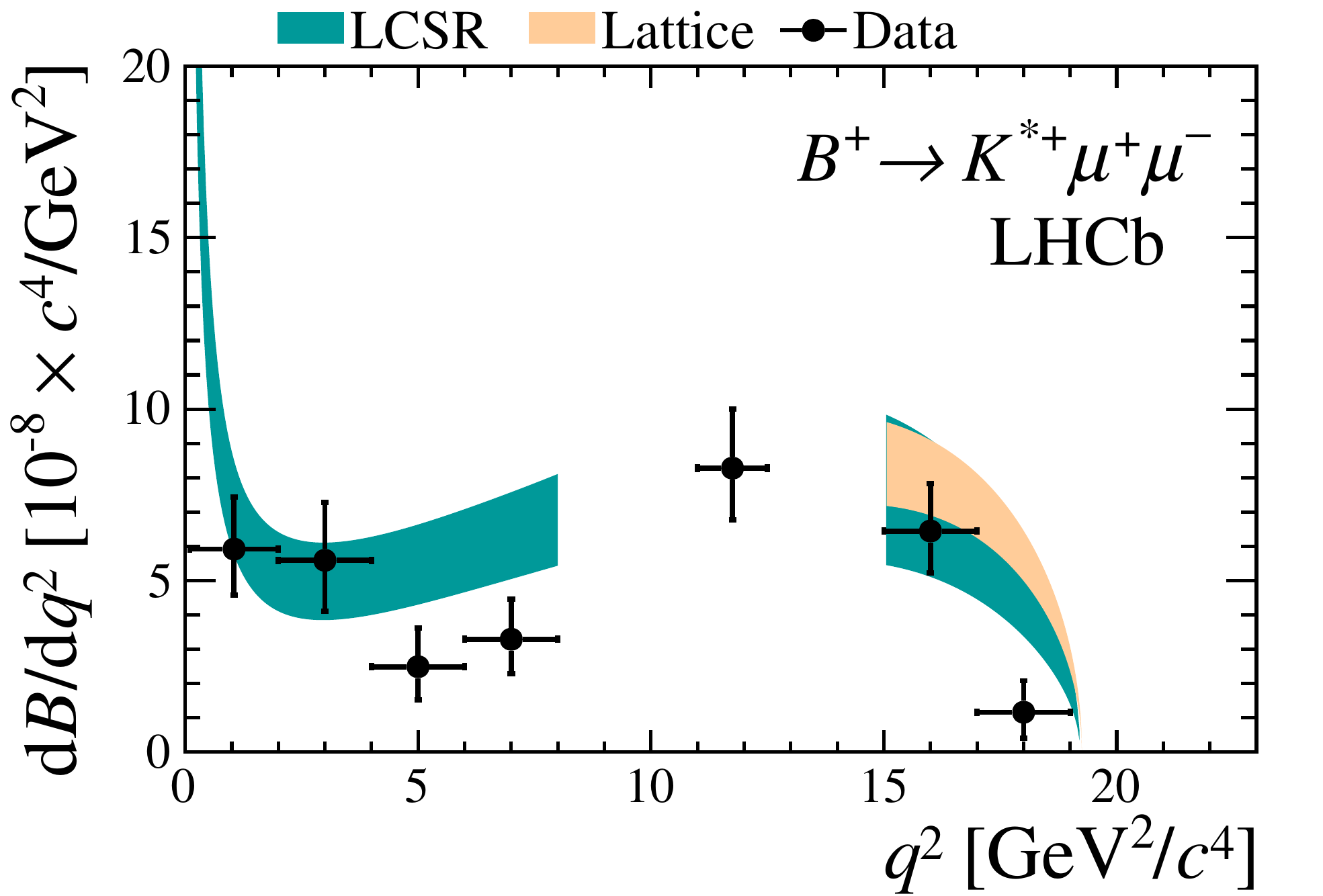}
  \caption{
    Differential branching fraction of (left) $\decay{\Bu}{\Kp\mumu}$ (middle) $\decay{\Bd}{K^0\mumu}$ and (right) $\decay{\Bu}{\Kstarp\mumu}$, overlaid with SM predictions~\protect\cite{Bobeth:2011gi,Bobeth:2011nj}.\label{fig:bfs}}
\end{figure}

Compared to angular observables, 
branching fraction measurements of $b\to s\mumu$ processes tend to have larger associated theory uncertainties, 
since they are directly impacted by the hadronic form-factors. 
However, the impact of theory uncertainties can be mitigated, by performing measurements of ratios of branching fractions where
form-factor uncertainties cancel at leading order. 
Examples of such quantities are the isospin asymmetry $A_I$ and the \CP-asymmetry $A_{\rm CP}$, defined as
\begin{eqnarray}
  A_I&=&\frac{\Gamma(\decay{\Bd}{K^{(*)0}\mumu})-\Gamma(\decay{\Bu}{K^{(*)+}\mumu})}{\Gamma(\decay{\Bd}{K^{(*)0}\mumu})+\Gamma(\decay{\Bu}{K^{(*)+}\mumu})},\\
  A_{\rm CP}&=& \frac{\Gamma(\bar{B}\to\bar{K}^{(*)}\mup\mun) - \Gamma(B\to K^{(*)}\mup\mun)}{\Gamma(\bar{B}\to\bar{K}^{(*)}\mup\mun) + \Gamma(B\to K^{(*)}\mup\mun)}.
\end{eqnarray}
In Refs.~\cite{Aaij:2014pli,Aaij:2014bsa}, $A_I$ and $A_{\rm CP}$ are found to be compatible with SM predictions~\cite{Lyon:2013gba,Altmannshofer:2008dz}. 
The corresponding differential branching fraction measurements for
the rare decays $\decay{\Bu}{\Kp\mumu}$, $\decay{\Bd}{K^0\mumu}$ and $\decay{\Bu}{\Kstarp\mumu}$ are given in Fig.~\ref{fig:bfs}.
They are compatible with, but tend to lie below, SM predictions~\cite{Bobeth:2011gi,Bobeth:2011nj}. 

Using $1\invfb$ of data taken during 2011, LHCb also determines the differential branching fractions for the rare decays $\decay{\Bd}{\Kstarz\mumu}$ and $\decay{\Bs}{\phi\mumu}$~\cite{Aaij:2013iag,Aaij:2013aln}.
The differential branching fractions tend to be below SM predictions both at low $q^2$, 
where updated light cone sum rule calculations are available~\cite{Straub:2015ica},
and at high $q^2$, where lattice calculations exist~\cite{Horgan:2013hoa,Horgan:2013pva,Horgan:2015vla}.
For the decay $\decay{\Bs}{\phi\mumu}$ the tension in the region $1<q^2<6\gevgevcccc$ corresponds to $3.1\,\sigma$. 
It is interesting to note, that the deviation of the branching fractions points to a deviation of the $b\to s\mumu$ couplings which is compatible with,
but less significant than, what is observed from the angular observables in $\decay{\Bd}{\Kstarz\mumu}$ at low $q^2$~\cite{Altmannshofer:2014rta,Horgan:2015vla}. 
Updated measurements of ${\cal B}(\decay{\Bd}{\Kstarz\mumu})$ and ${\cal B}(\decay{\Bs}{\phi\mumu})$
using the full Run I data sample are currently in preparation to clarify the situation. 

\section{Branching fraction of $\decay{B_{(s)}^0}{\pip\pim\mumu}$}
The $\pip\pim\mumu$ final state can be reached from both the decay of a $\Bd$ meson and the decay of a $\Bs$ meson.
The $\Bd$ decay is expected to be dominated by the $b\to d\mumu$ transition $\decay{\Bd}{\rhoz\mumu}$,
the $\Bs$ decay by the $b\to s\mumu$ transition $\decay{\Bs}{f^0(980)\mumu}$.
While $b\to d$ decays 
are expected to be suppressed by the factor $|V_{td}|^2/|V_{ts}|^2\sim 0.04$ 
compared to $b\to s$ transitions in the SM,
this is not necessarily the case for SM extensions. 

\begin{figure}
  \centering
  \includegraphics[width=0.28\linewidth]{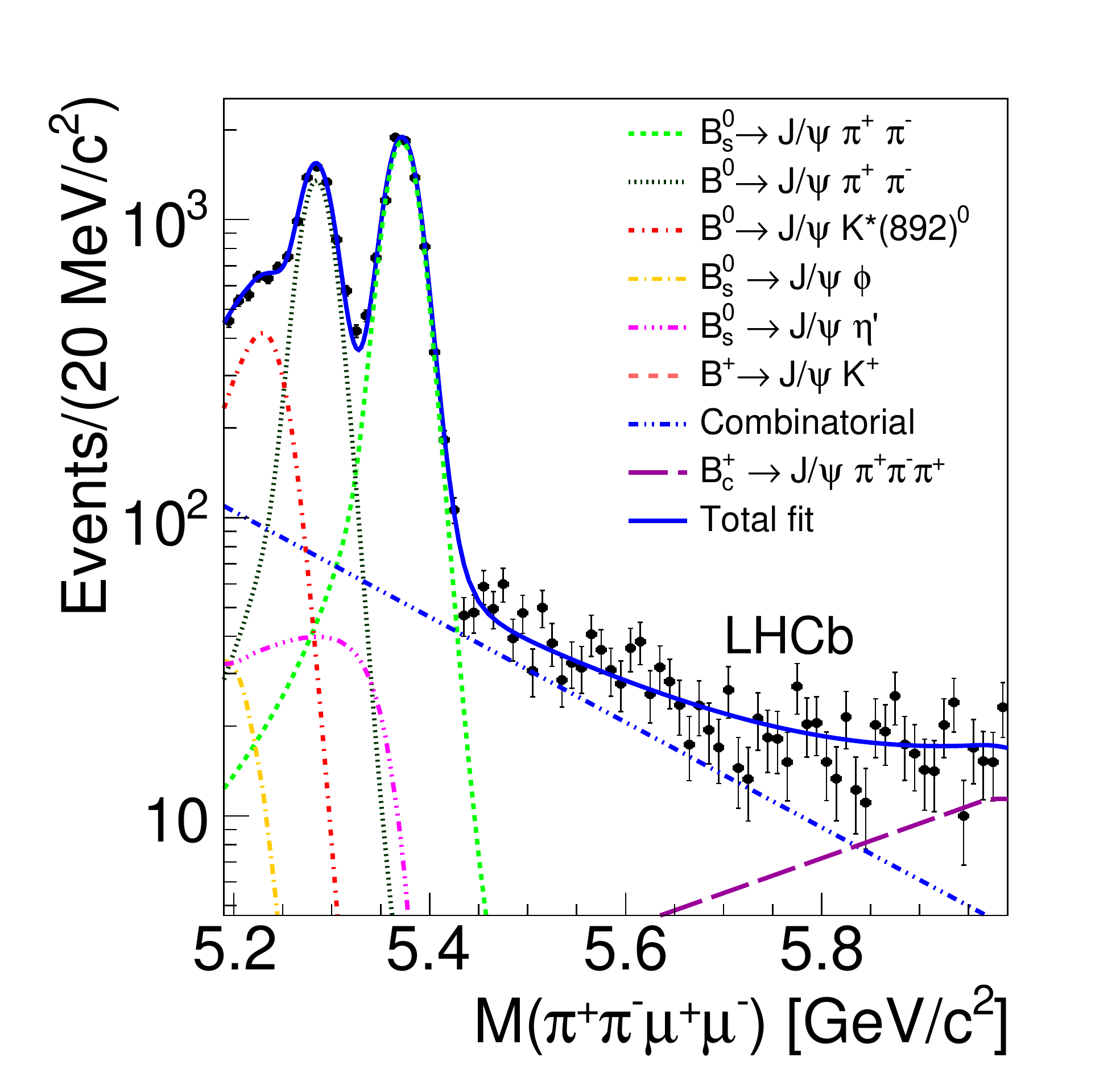} 
  \includegraphics[width=0.28\linewidth]{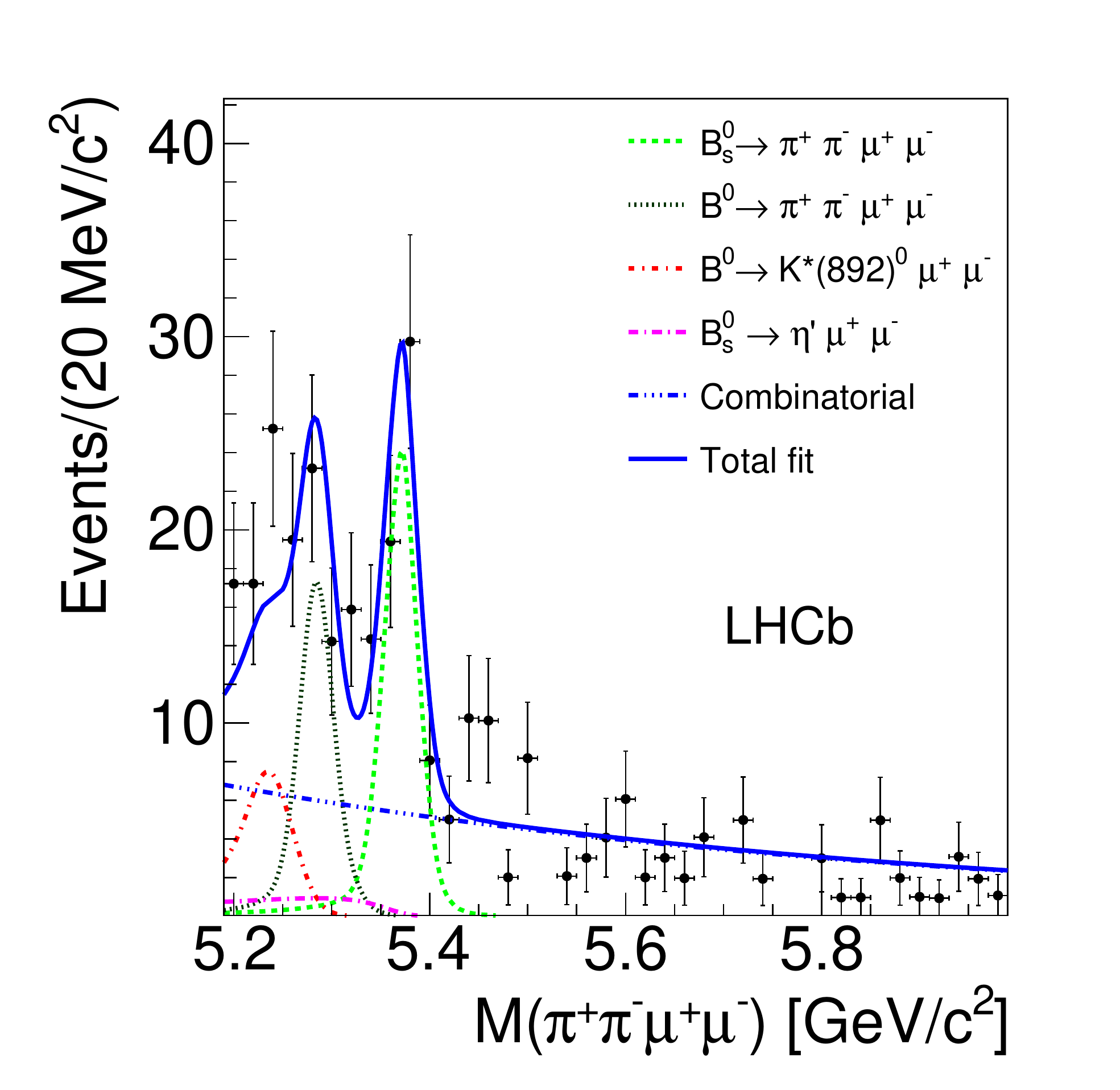}
  \caption{Invariant mass of the $\pip\pim\mup\mun$ final state for (left) tree-level charmonium decays $\decay{\Bd}{\jpsi\pip\pim}$ and $\decay{\Bs}{\jpsi\pip\pim}$ and (right) the rare decays $\decay{\Bd}{\pip\pim\mumu}$ and $\decay{\Bs}{\pip\pim\mumu}$.\label{fig:pipimumu}}
\end{figure}

The $\pip\pim\mup\mun$ final state is studied using the full Run I data sample taken by the LHCb experiment~\cite{Aaij:2014lba},
corresponding to an integrated luminosity of $3\invfb$. 
The invariant mass of the $\pip\pim$ system is required to be in the range $0.5-1.3\gevcc$ containing both the $\rhoz$ as well as the $f^0(980)$ resonance. 
Figure~\ref{fig:pipimumu} gives the invariant mass distribution of the $\pip\pim\mumu$ system for the charmonium modes $\decay{B^0_{(s)}}{\jpsi\pip\pim}$,
that are used as control decays for the fit model,
as well as the signal decays $\decay{B^0_{(s)}}{\pip\pim\mumu}$. 
The signal yields are found to be $40\pm10\pm3$ for the $\decay{\Bd}{\pip\pim\mumu}$ decay and
$55\pm10\pm5$ for the $\decay{\Bs}{\pip\pim\mumu}$ decay,
resulting in significances of $4.8\,\sigma$ and $7.2\,\sigma$, respectively. 
The branching fractions are determined with respect to the normalisation mode $\decay{\Bd}{\jpsi\Kstarz}$. 
They are found to be
\begin{eqnarray*}
{\cal B}(\Bs\to\pip\pim\mup\mun)&=&(8.6\pm 1.5_{\rm stat.}\pm0.7_{\rm syst.}\pm0.7_{\rm norm.})\times 10^{-8},\\
{\cal B}(\Bd\to\pip\pim\mup\mun)&=&(2.11\pm 0.51_{\rm stat.}\pm0.15_{\rm syst.}\pm0.16_{\rm norm.})\times 10^{-8},
\end{eqnarray*}
when correcting for the $q^2$ regions removed by the vetoes of the $\jpsi\pip\pim$ and $\psitwos\pip\pim$ decays, 
in agreement with SM predictions~\cite{Kruger:1997jk,Beneke:2004dp,Li:2008tk,Colangelo:2010bg}. 

\section{The rare baryonic decay $\decay{\Lb}{\Lambda\mumu}$}
The study of the rare $\Lb$ decay $\decay{\Lb}{\Lambda\mumu}$ is of particular interest due to the half-integer spin of the $\Lb$ baryon and 
the hadronic dynamics involving the heavy $b$ and a light diquark system.
Furthermore, the $\Lambda$ decays weakly into the $p\pim$ final state,
allowing access to new and complementary information compared to mesonic $b\to s\mumu$ decays~\cite{Boer:2014kda}. 

\begin{figure}
  \centering
  \includegraphics[width=0.32\linewidth]{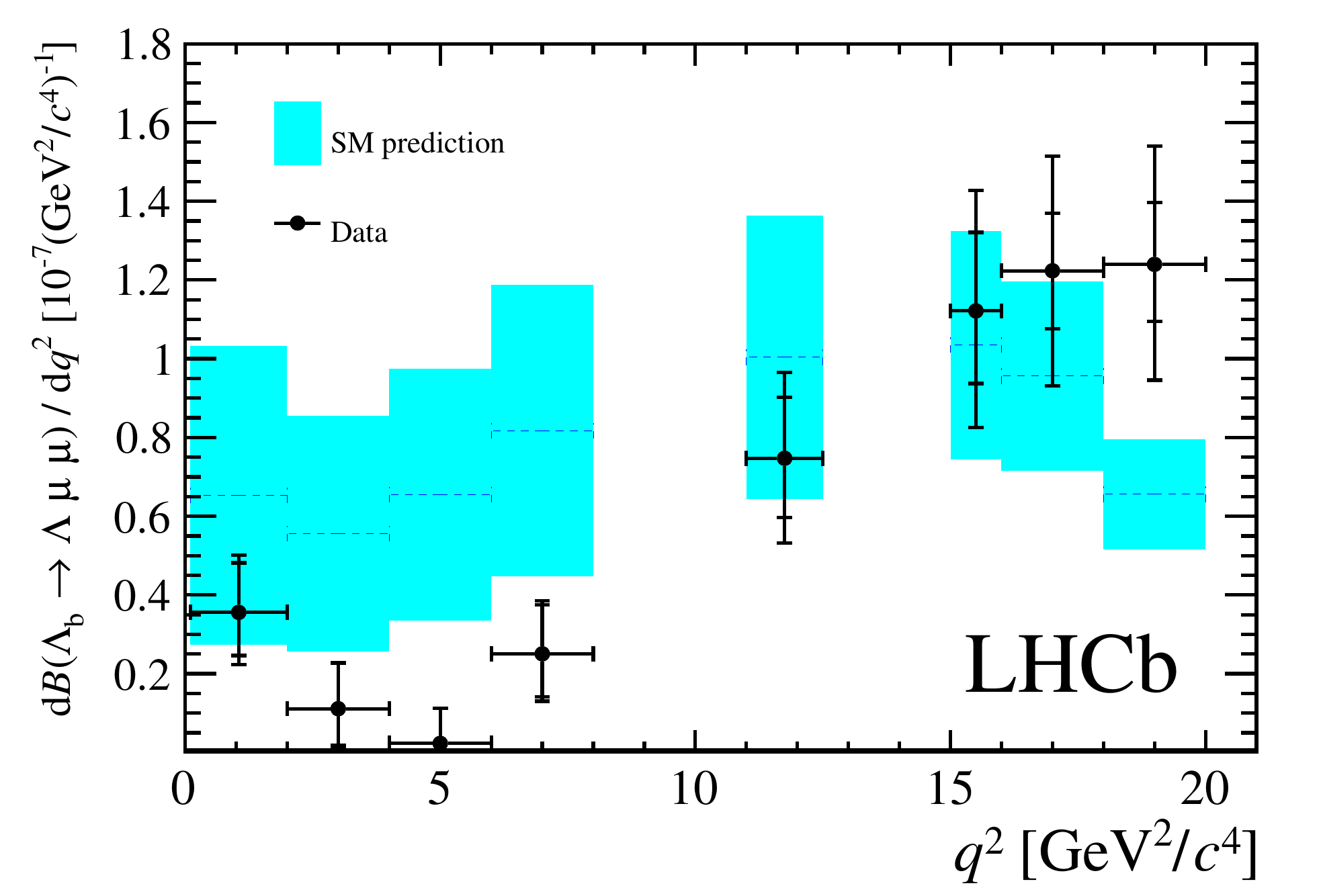} 
  \includegraphics[width=0.32\linewidth]{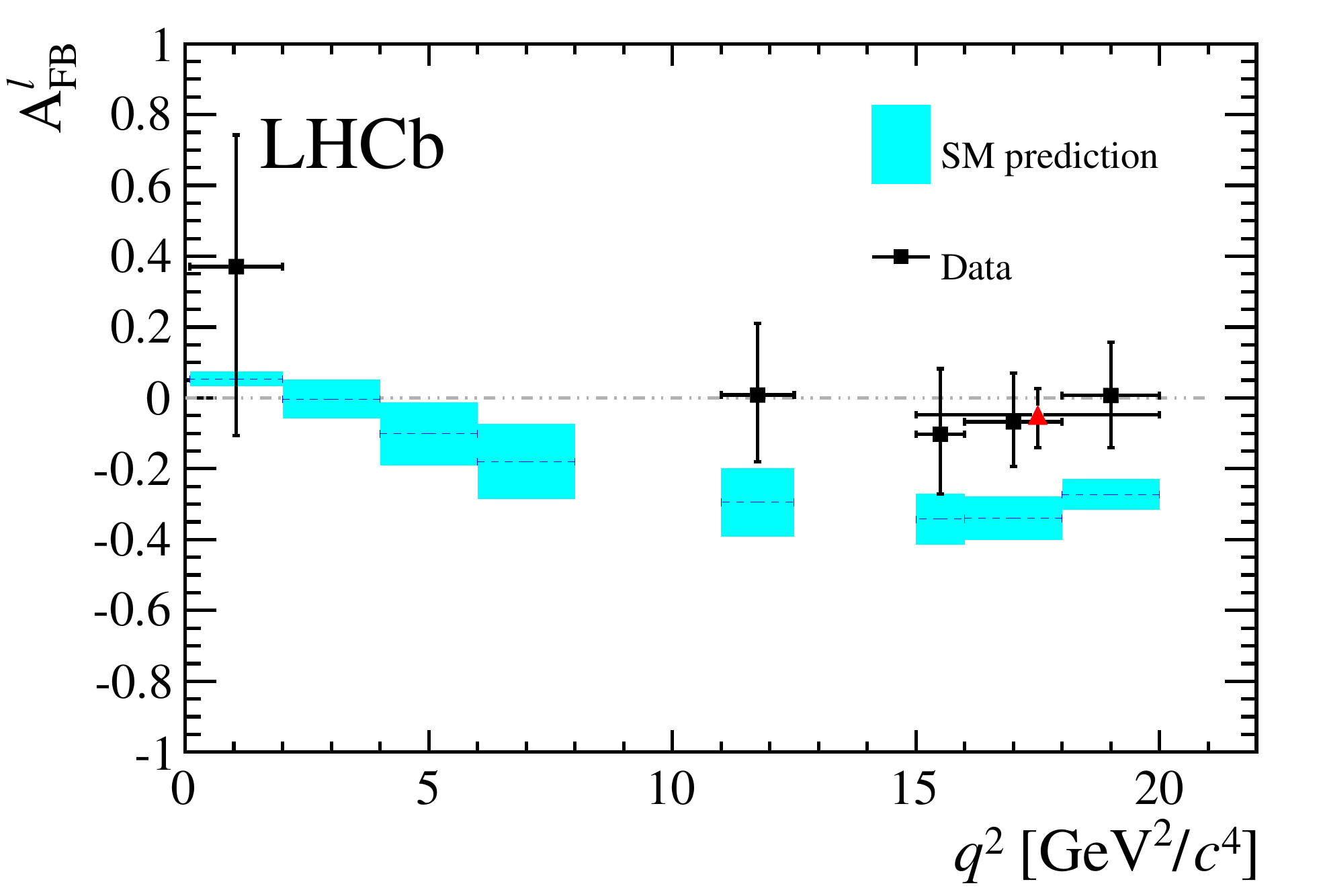}
  \includegraphics[width=0.32\linewidth]{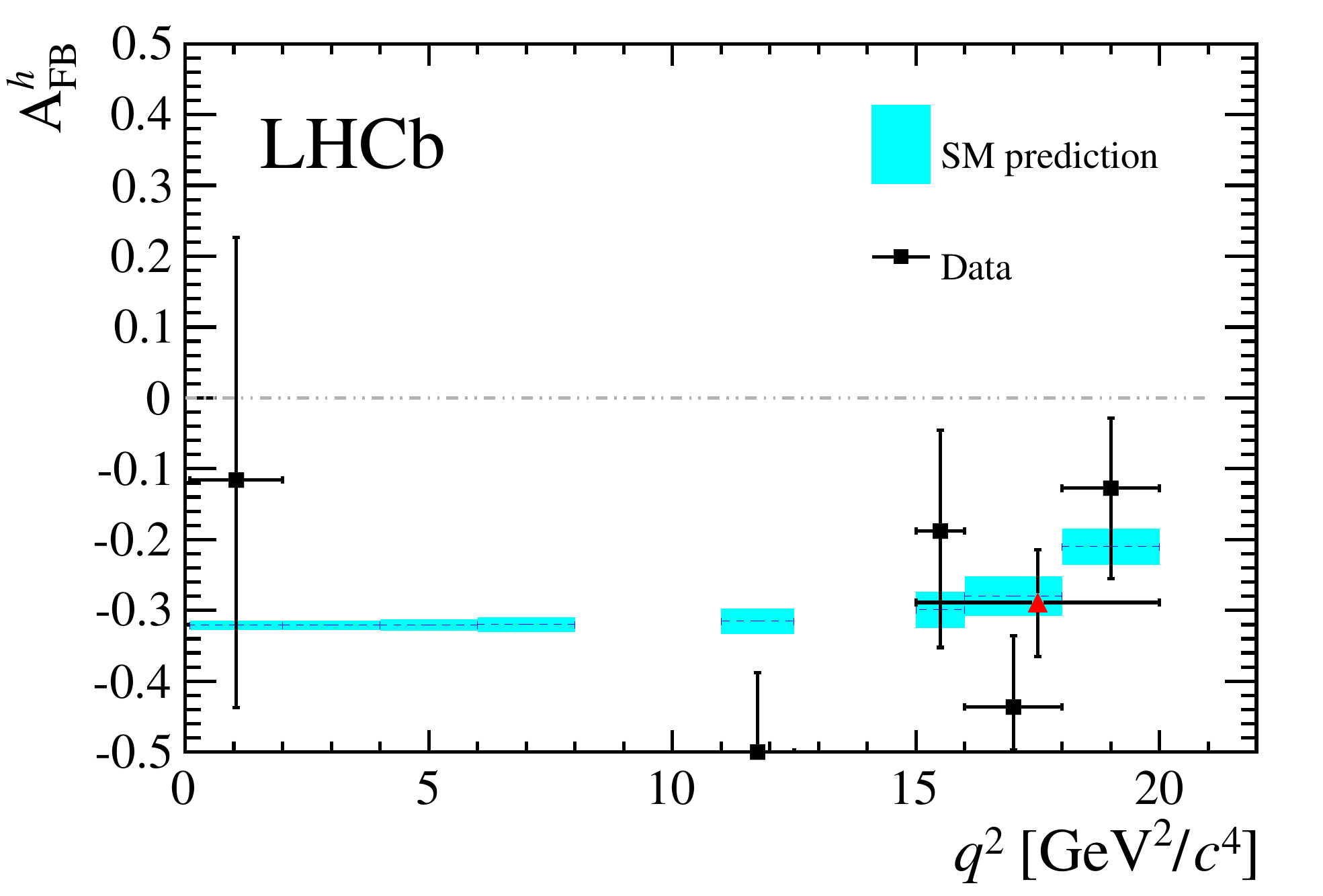}
  \caption{The (left) differential branching fraction and (middle) leptonic, as well as the (right) hadronic forward-backward asymmetry,
    overlaid with SM predictions~\protect\cite{Detmold:2012vy,Meinel:2014wua}.\label{fig:lambdamumu}}
\end{figure}

The decay was previously studied in~\cite{Aaltonen:2011qs,Aaij:2013hna}, where no evidence for signal in the $q^2$ region below the $\jpsi$ was found. 
An updated analysis is performed, using the full LHCb Run I data sample~\cite{Aaij:2015xza}. 
Figure~\ref{fig:lambdamumu} gives the differential branching fraction. 
Evidence for signal is found below the charmonium resonances at low $q^2$,
the differential branching fraction for the high $q^2$ range $15<q^2<20\gevgevcccc$
is determined to be $(1.18^{+0.09}_{-0.08}\pm0.03\pm0.27)\times 10^{-7}\gev^{-2}c^{4}$~\cite{Aaij:2015xza}. 
Angular analyses are performed for the $q^2$ bins where evidence for signal is found  
and the angular observables $A_{\rm FB}^\ell$ and $A_{\rm FB}^h$,
the forward-backward asymmetries in the dimuon and hadron system, are determined. 
As shown in Fig.~\ref{fig:lambdamumu}, $A_{\rm FB}^\ell$ and $A_{\rm FB}^h$ are 
found to be in good agreement with SM predictions~\cite{Meinel:2014wua,Detmold:2012vy}. 

\section{A test of lepton universality using the decay $\decay{\Bu}{\Kp e^+e^-}$}
The ratio $R_K$ in the $q^2$ region $[q^2_{\rm min},q^2_{\rm max}]$ is defined as
\begin{eqnarray}
  R_K &=& \frac{\int_{q^2_{\rm min}}^{q^2_{\rm max}}\frac{{\rm d}\Gamma[\Bu\to\Kp\mup\mun]}{{\rm d}q^2}{\rm d}q^2}{\int_{q^2_{\rm min}}^{q^2_{\rm max}}\frac{{\rm d}\Gamma[\Bu\to\Kp e^+e^-]}{{\rm d}q^2}{\rm d}q^2}, 
\end{eqnarray}
where $\Gamma$ denotes the $q^2$ dependent partial width. 
Due to the universal coupling of $\gamma$ and $Z_0$ bosons to leptons, 
$R_K$ in the region $1<q^2<6\gevgevcccc$ is predicted to be one with an uncertainty of less than $10^{-3}$~\cite{Hiller:2003js,Bobeth:2007dw}. 
Small corrections to the ratio arise only from phase-space effects and Higgs penguin contributions. 

\begin{figure}
  \centering
  \includegraphics[width=0.32\linewidth]{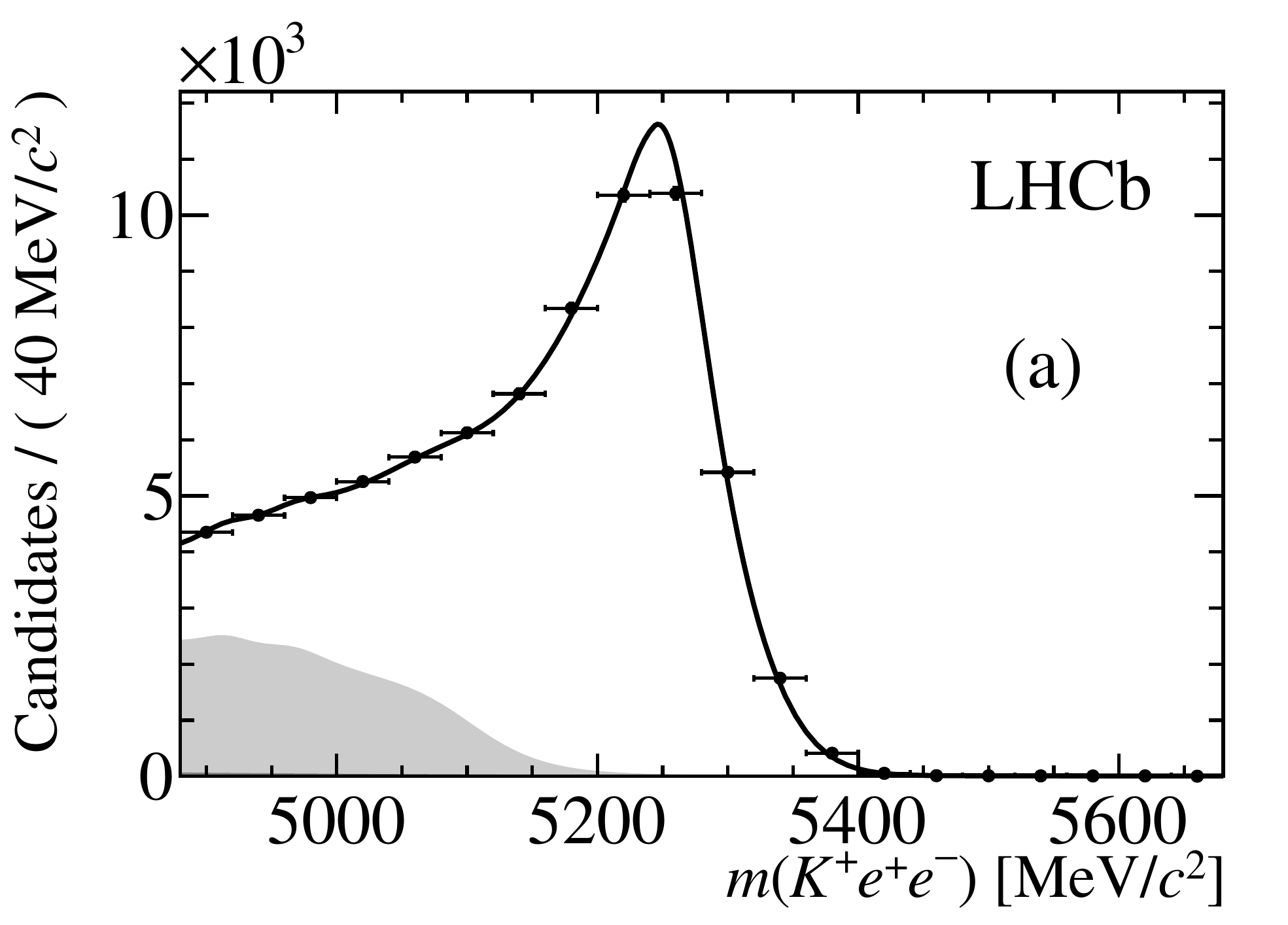} 
  \includegraphics[width=0.32\linewidth]{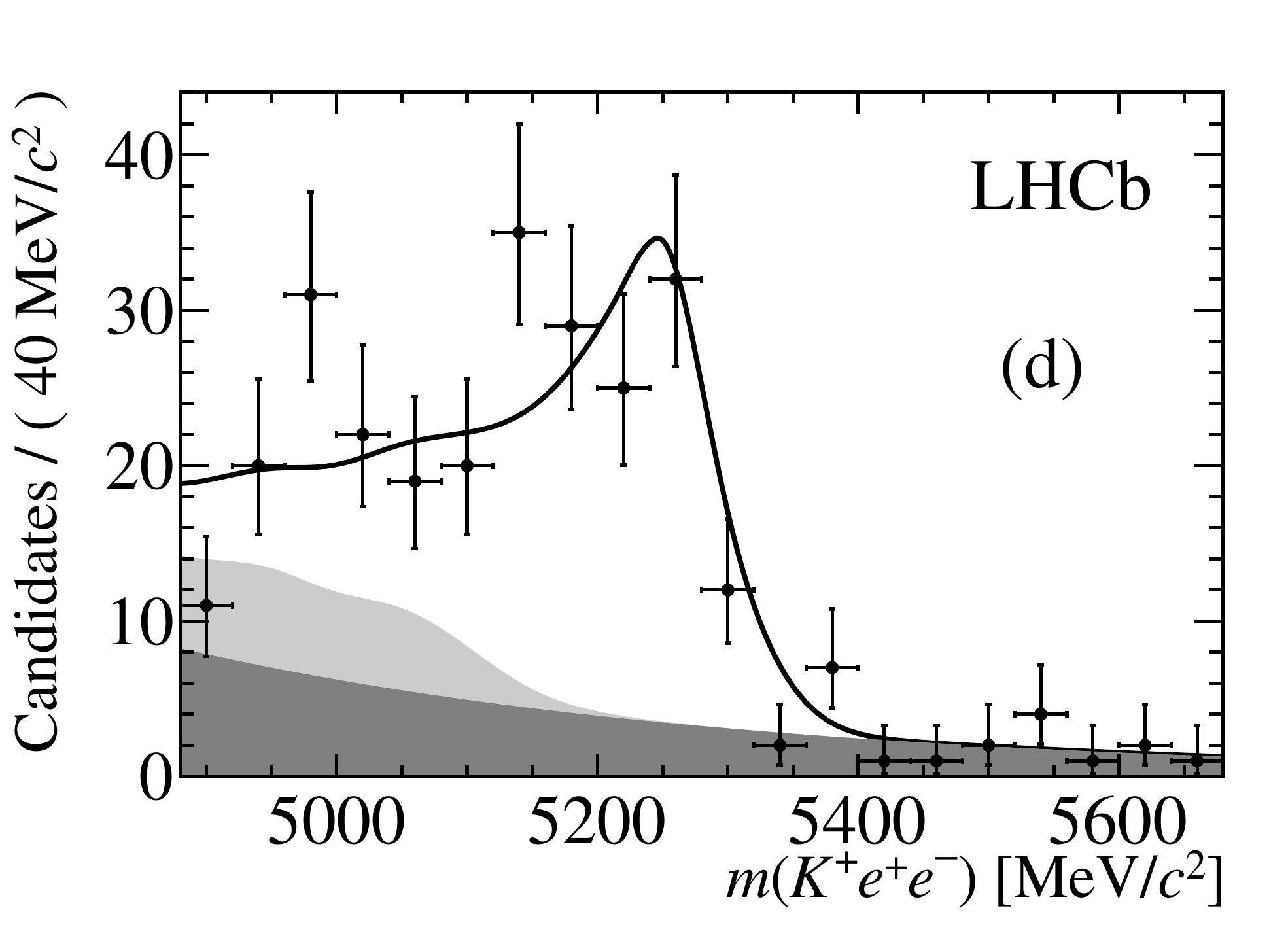} 
  \includegraphics[width=0.32\linewidth]{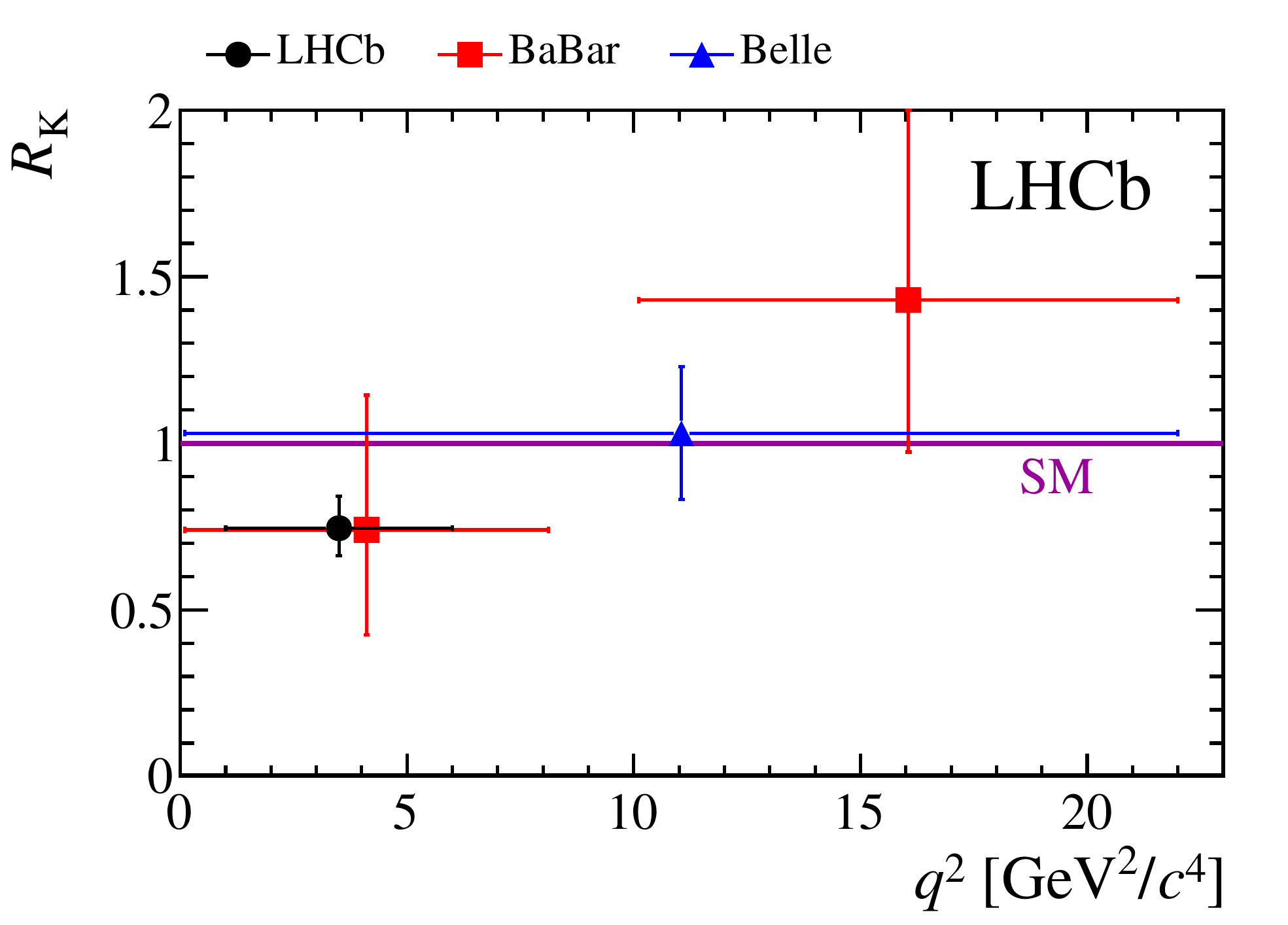} 
  \caption{
    (Left) $\decay{\Bu}{\jpsi\Kp}$ and (middle) $\decay{\Bu}{\Kp e^+e^-}$ signal candidates, triggered on the electron in the final state. 
    (Right) The ratio $R_K$ as determined by LHCb~\protect\cite{Aaij:2014ora}, BaBar~\protect\cite{Lees:2012tva} and Belle~\protect\cite{Wei:2009zv} for different $q^2$ ranges.\label{fig:rk}}
\end{figure}

The measurement is experimentally challenging due to a lower trigger efficiency for electrons compared to muons and
the higher emission of Bremsstrahlung which deteriorates the resolution of the invariant mass of the $\Kp\ep\en$ system. 
In the range $1<q^2<6\gevgevcccc$, $R_K$ is determined to be
\begin{eqnarray*}
R_K &=& 0.745^{+0.090}_{-0.074}{\rm(stat.)}\pm0.036{\rm(syst.)},
\end{eqnarray*}
which corresponds to a deviation of $2.6\,\sigma$ from the SM prediction~\cite{Aaij:2014ora}. 
Figure~\ref{fig:rk} shows the decay $\decay{\Bu}{\jpsi\Kp}$,
which is used to study the effect of Bremsstrahlung and understand the relative efficiency between reconstructing dimuon and dielectron modes. 
The signal decay $\decay{\Bu}{\Kp e^+e^-}$ is also given, 
as well as the LHCb measurement of $R_K$~\cite{Aaij:2014ora} in comparison with results from the B factories~\cite{Lees:2012tva,Wei:2009zv}. 
Since $R_K$ is free from hadronic uncertainties, 
the result received considerable attention from theory~\cite{Glashow:2014iga,Crivellin:2015mga,Hiller:2014yaa,Biswas:2014gga,Gripaios:2014tna,Sahoo:2015wya,Ghosh:2014awa,Hurth:2014vma}. 
Further tests of lepton universality are in preparation, 
including the measurements of $R_{K^*}$ and $R_\phi$. 

\section{Angular analysis of $\decay{\Bd}{\Kstarz e^+e^-}$}
The study of rare decays with electrons in the final state allows to perform analyses at very low $q^2$, due to the tiny electron mass.
At low $q^2$, the contribution from Feynman diagrams in which a virtual photon couples to the lepton pair dominates. 
This allows to probe the photon polarisation, which is left-handed in the SM. 

\begin{figure}
  \centering
  \includegraphics[width=0.32\linewidth]{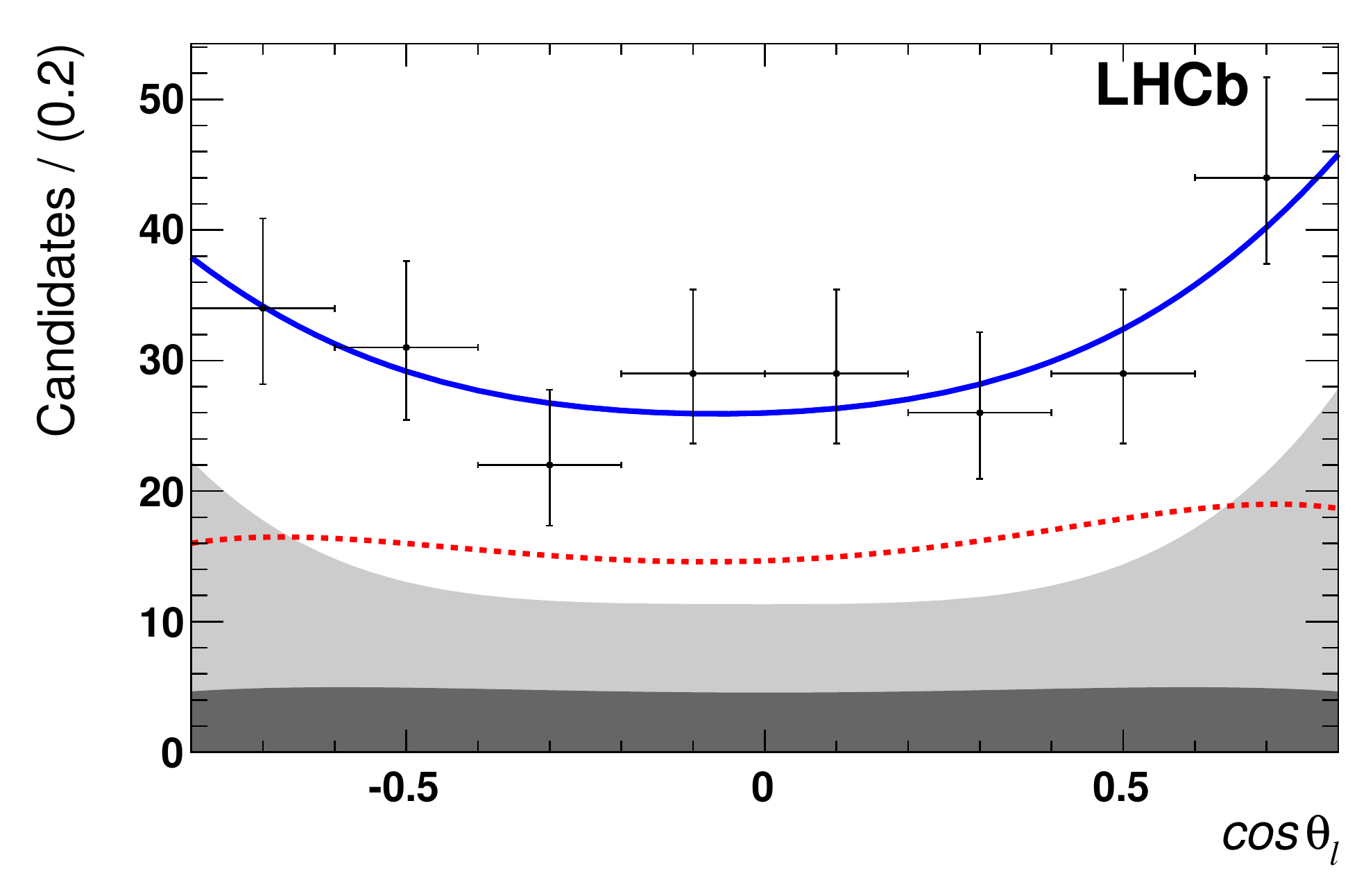}
  \includegraphics[width=0.32\linewidth]{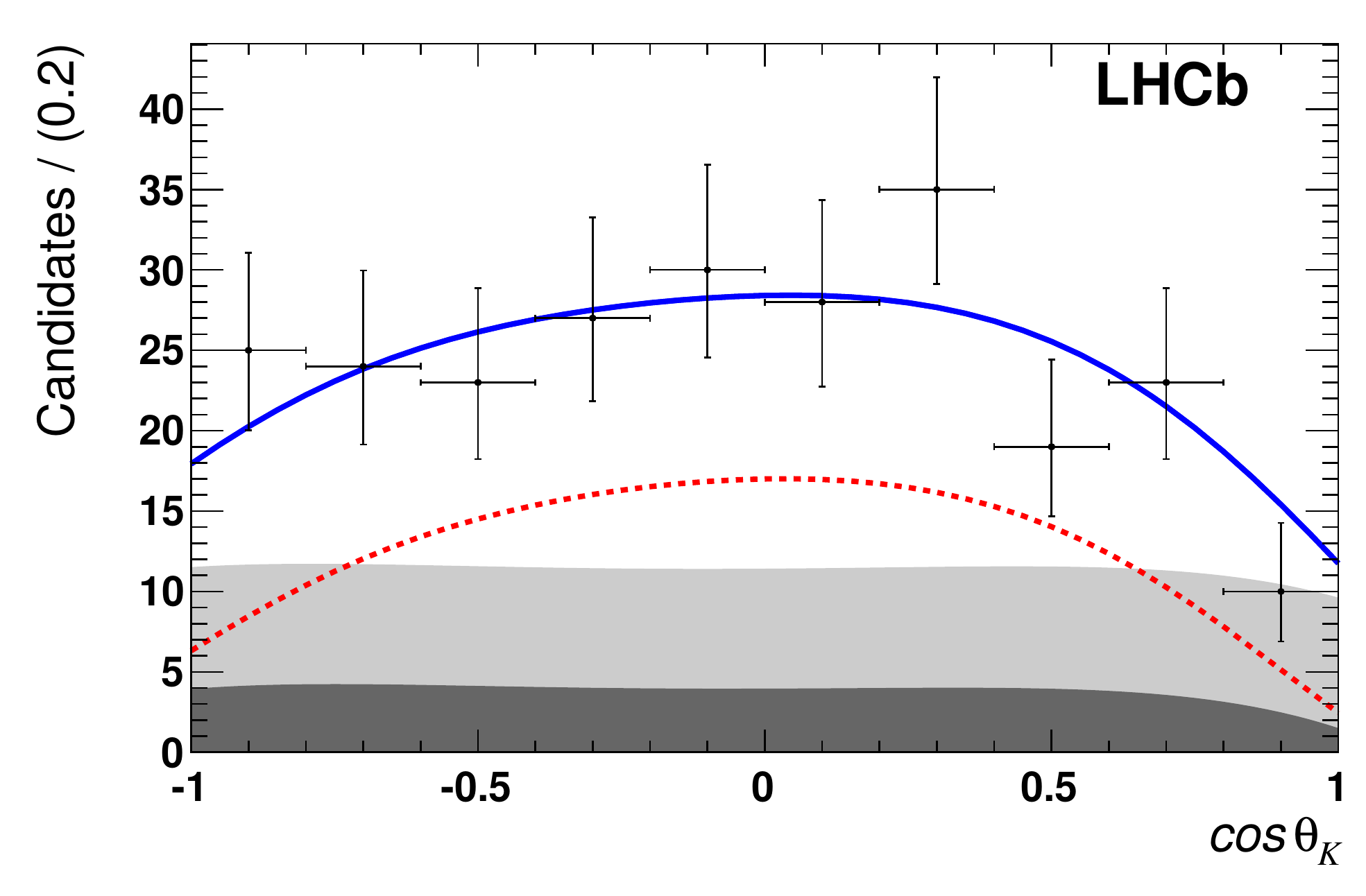}
  \includegraphics[width=0.32\linewidth]{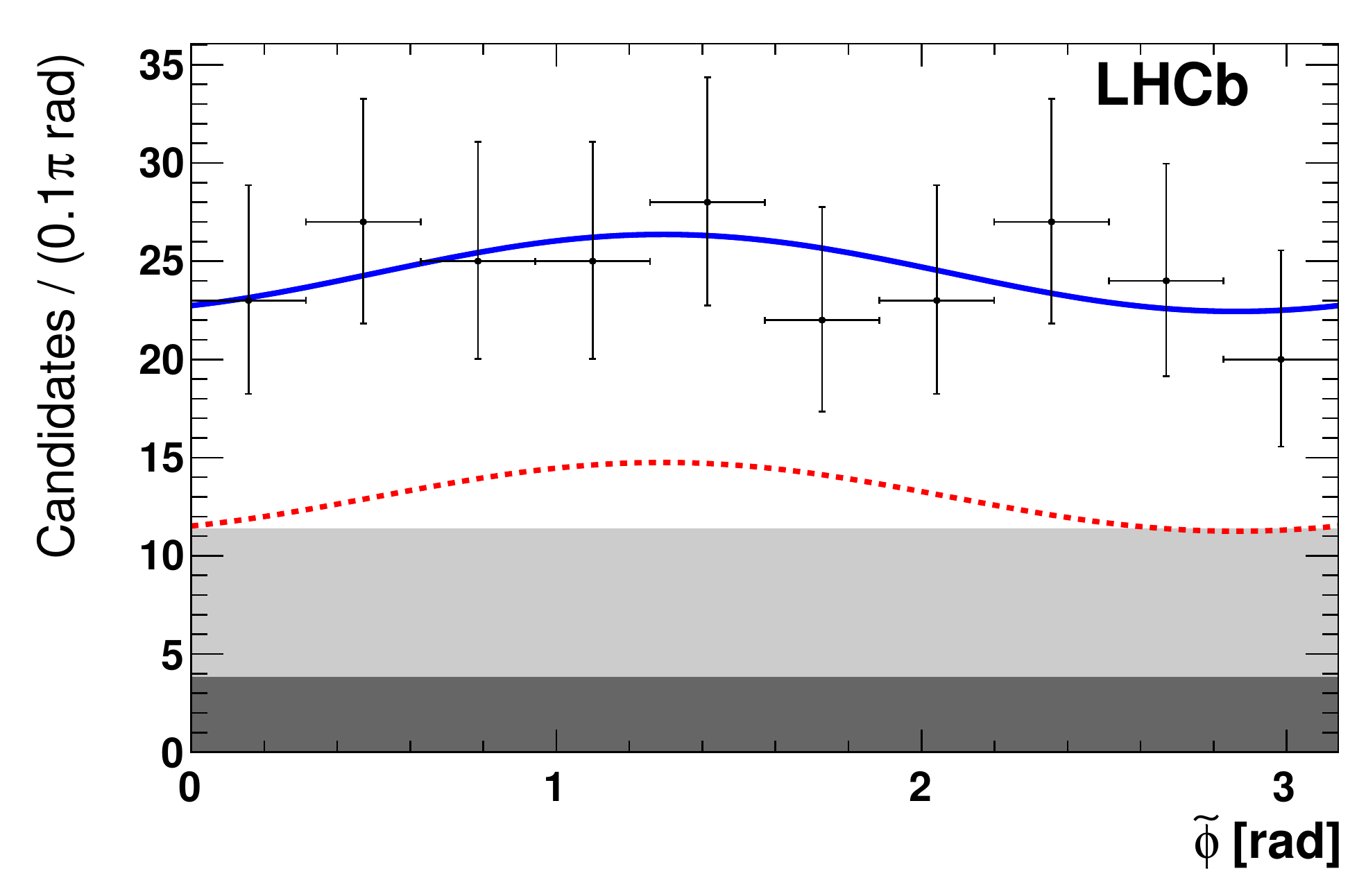}
  \caption{The three decay angles (left) $\cos\thetal$ (middle) $\cos\thetak$ and (right) $\phi$ of $\decay{\Bd}{\Kstarz\ep\en}$ signal candidates, overlaid with the projections of the fitted probability density function.\label{fig:kstaree}}
\end{figure}

LHCb performs an angular analysis of the decay $\decay{\Bd}{\Kstarz\ep\en}$ in the $q^2$ range $0.002<q^2<1.120\gevgevcccc$~\cite{Aaij:2015dea}. 
The four angular observables $F_{\rm L}$, $A_{\rm T}^{(2)}$, $A_{\rm T}^{\rm Re}$ and $A_{\rm T}^{\rm Im}$
are determined from an unbinned maximum likelihood fit to the decay angles $\cos\thetal$, $\cos\thetak$ and $\phi$.
Of particular interest are the observables $A_{\rm T}^{(2)}$ and $A_{\rm T}^{\rm Im}$ that are sensitive to the photon polarization. 
Figure~\ref{fig:kstaree} gives the angular fit projections. 
The measured angular observables are
\begin{eqnarray*}
F_{\rm L} &=& +0.16\pm0.06\pm0.03\\
A_{\rm T}^{(2)} &=& -0.23\pm0.23\pm0.05\\
A_{\rm T}^{\rm Re} &=& +0.10\pm0.18\pm0.05\\
A_{\rm T}^{\rm Im} &=& +0.14\pm0.22\pm0.05,
\end{eqnarray*}
which is in good agreement with SM predictions~\cite{Becirevic:2011bp,Jager:2012uw}.
The constraints from $A_{\rm T}^{(2)}$, $A_{\rm T}^{\rm Im}$ on the contributions
from right-handed currents are more precise than those obtained
from the average of the time dependent \CP-asymmetries in radiative $\decay{\Bd}{\Kstarz(\to\KS\piz)\gamma}$ decays~\cite{Aubert:2008gy,Ushiroda:2006fi}. 

\section{Conclusions}
Most of the observables in rare decays are found to be in good agreement with SM predictions.
However, three interesting tensions emerge:
An update of the angular analysis of the decay $\decay{\Bd}{\Kstarz\mumu}$ confirms a deviation of the angular observable $P_5^\prime$ in the two $q^2$ bins
$4<q^2<6\gevgevcccc$ and $6<q^2<8\gevgevcccc$, with a significance of $2.9\,\sigma$ in each; 
Furthermore, the branching fraction of the rare decay $\decay{\Bs}{\phi\mumu}$
in the range $1<q^2<6\gevgevcccc$
is $3.1\,\sigma$ lower than a recently updated theory prediction; 
Finally, the measurement of $R_K$ shows a tension with lepton universality at $2.6\,\sigma$. 

Consistent NP explanations of all observed tensions in rare decays exist,
and first global fits including the updated results on $\decay{\Bd}{\Kstarz\mumu}$
angular observables 
prefer the NP solution over the SM by $3.7\,\sigma$~\cite{Altmannshofer:2015sma}. 
However, it is too early to speak of clear signs of processes beyond the SM; 
Unexpectedly large hadronic contributions still can not be excluded. 
The results clearly motivate future work, both from theory, 
as well as from experiment, 
where complementary measurements of rare $b\to (s,d)\ell\ell$ processes will be performed. 
For the Run I LHCb data, this includes an update of the analysis of the decay $\decay{\Bs}{\phi\mumu}$ 
and an updated branching fraction measurement of the decay $\decay{\Bd}{\Kstarz\mumu}$. 
In addition, further tests of lepton universality and lepton number violation are in preparation. 
The data sample LHCb will collect during Run II will 
further improve the experimental sensitivity 
and allow to probe the operator structure of rare decays with unprecedented precision.

\end{document}